\documentclass[journal]{IEEEtran}

\usepackage{amsmath} 
\usepackage{amssymb}
\usepackage{booktabs}
\usepackage{subcaption}
\usepackage{hyperref}
\usepackage{colortbl}
\usepackage{xcolor}
\usepackage{multirow}
\usepackage{pgfplots}
\pgfplotsset{compat=1.17}
\usepackage{todonotes}

\newcommand{\applycolormap}[1]{%
    \pgfmathsetmacro{\value}{#1}%
    \ifdim \value pt > 19 pt \cellcolor[HTML]{A50026}% Dark red (highest error)
    \else\ifdim \value pt > 18 pt \cellcolor[HTML]{D73027}%
    \else\ifdim \value pt > 17 pt \cellcolor[HTML]{F46D43}%
    \else\ifdim \value pt > 16 pt \cellcolor[HTML]{FDAE61}%
    \else\ifdim \value pt > 15 pt \cellcolor[HTML]{FEE08B}%
    \else\ifdim \value pt > 14 pt \cellcolor[HTML]{FFFFBF}%
    \else\ifdim \value pt > 13 pt \cellcolor[HTML]{D9EF8B}%
    \else\ifdim \value pt > 12 pt \cellcolor[HTML]{A6D96A}%
    \else\ifdim \value pt > 11 pt \cellcolor[HTML]{66BD63}%
    \else\ifdim \value pt > 9 pt \cellcolor[HTML]{1A9850}%
    \else\cellcolor[HTML]{006837}% Dark green (lowest error)
    \fi\fi\fi\fi\fi\fi\fi\fi\fi\fi
    #1
}

\begin{document}

\title{Advancing Brainwave-Based Biometrics: \\ A Large-Scale, Multi-Session Evaluation}

\author{
    \IEEEauthorblockN{Matin Fallahi\IEEEauthorrefmark{1}, Patricia Arias-Cabarcos\IEEEauthorrefmark{4}\textsuperscript{a}\thanks{\textsuperscript{a}Current Affiliation: European Commission, Joint Research Centre (JRC), Ispra, Italy.}, Thorsten Strufe\IEEEauthorrefmark{1}}
    \\
    \IEEEauthorblockA{\IEEEauthorrefmark{1}KASTEL, Karlsruhe Institute of Technology (KIT), Germany \\ 
    Email: \{matin.fallahi, strufe\}@kit.edu}
    \IEEEauthorblockA{\IEEEauthorrefmark{4}Paderborn University, Germany \\ 
    Email: pac@mail.upb.de}
}

% The paper headers
\markboth{Journal of \LaTeX\ Class Files,~Vol.~14, No.~8, August~2015}%
{Shell \MakeLowercase{\textit{et al.}}: Bare Demo of IEEEtran.cls for IEEE Journals}

% make the title area
\maketitle

\begin{abstract}
The field of brainwave-based biometrics has gained attention for its potential to revolutionize user authentication through hands-free interaction, resistance to shoulder surfing, continuous authentication, and revocability. However, current research often relies on single-session or limited-session datasets with fewer than 55 subjects, raising concerns about the generalizability of the findings. To address this gap, we conducted a large-scale study using a public brainwave dataset comprising 345 subjects and over 6,007 sessions (an average of 17 per subject) recorded over five years using three headsets. Our results reveal that deep learning approaches significantly outperform hand-crafted feature extraction methods. We also observe Equal Error Rates  (EER) increases over time (e.g., from 6.7\% after 1 day to 14.3\% after a year). Therefore, it is necessary to reinforce the enrollment set after successful login attempts. Moreover, we demonstrate that fewer brainwave measurement sensors can be used, with an acceptable increase in EER, which is necessary for transitioning from medical-grade to affordable consumer-grade devices. 
Finally, we compared our results to prior work and existing biometric standards. While our performance is on par with or exceeds previous approaches, it still falls short of industrial benchmarks. Based on the results, we hypothesize that further improvements are possible with larger training sets. To support future research, we have open-sourced our analysis code.
\end{abstract}


\IEEEpeerreviewmaketitle

\section{Introduction}
Authentication is the cornerstone of securing digital interactions. While passwords have long been the dominant method, biometrics are increasingly preferred. According to the 2024 FIDO Survey\footnote{\scriptsize\url{https://fidoalliance.org/wp-content/uploads/2024/10/Barometer-Report-2024-Oct-29.pdf}} of 10,000 consumers, 28\% preferred biometrics for signing into online accounts, apps, and smart devices, making it the most popular choice.  
Beyond established methods---such as fingerprint and face recognition---emerging biometric technologies are being explored to enhance usability and security~\cite{nath2023acute,raurale2021emg}. 
Brainwave authentication has gained attention for its hands-free operation, which enhances usability in devices like XR systems~\cite{lin2018brain, fallahi2024beyond}. 
It offers resistance to shoulder surfing by relying on distinct neural activity patterns rather than observable actions or physical characteristics~\cite{arias2023performance}. 
Additionally, brainwave-based systems can enable continuous authentication, improving security by monitoring user identity in active tracking~\cite{nakanishi2015brain}.
Furthermore, research indicates that in cases of brainwave sample leakage, changing the stimulus can effectively enable template renewability and revocability, thereby maintaining security~\cite{lin2018brain}. This contrasts with conventional biometric authentication, where compromised raw templates typically cannot be revoked or updated. However, brainwave authentication is still in its early stages, and both permanence and performance remain significant challenges.

Prior studies on brainwave authentication mainly rely on limited public datasets or small, privately collected datasets. 
Consequently, they either focus on single-session datasets~\cite{schons2018convolutional, bidgoly2022towards, arias2023performance,fallahi2023brainnet} or small-scale multi-session datasets~\cite{das2016eeg,wu2018eeg,maiorana2021learning,debie2021session}. While single-session studies completely overlook session and temporal effects~\cite{chaurasia2024neuroidbench, huang2022m3cv}, multi-session studies remain limited to a maximum of 55 subjects~\cite{debie2021session} and 270 sessions in total~\cite{maiorana2017longitudinal} (six sessions per subject) so far. These limitations increase the risk of overfitting recognition models and limit the generalizability of results, making it difficult to identify key factors that affect long-term reliability. Importantly, existing multi-session studies rely on a single EEG device, which prevents assessment of hardware variability, although users may access the same profile from different devices in real-world settings. Moreover, the lack of publicly accessible code and data hinders reproducibility and further development, in contrast to the face recognition field, which benefits from open data and collaborative research.

To address this gap and better understand how different parameters affect the performance of brainwave authentication over time, we analyzed the recently published PEERS dataset~\cite{kahana2024penn}, originally designed for large-scale, longitudinal studies of memory-related electrophysiology. The dataset includes recordings from 345 subjects and 6,007 sessions collected over five years using three different headsets. This dataset contains more than six times the number of subjects and 22 times the number of sessions compared to previous brainwave authentication evaluations. Our contributions are:

\begin{itemize}
    \item \textbf{Benchmark (Sec \ref{sec:benchmarke}):} To identify the best pipeline for multi-session brainwave authentication, we conducted a comprehensive evaluation of hand-crafted and deep learning-based feature extraction methods. The results showed that deep learning methods reduced the error rate by up to two times.

    \item  \textbf{Analysis (Sec \ref{sec:analysis})}: To better understand the parameters influencing performance, we investigated the effects of test size, session intervals, and device type. Our analysis revealed several key findings. First, a larger number of subjects in the evaluation set does not necessarily increase authentication error, while smaller datasets lead to higher uncertainty due to limited observations. Second, authentication errors increase over time by 100\% after one year. This suggests that updating the enrollment dataset after successful logins may help reduce long-term error rates. Third, visualization of the embedding space shows that samples from the same session and recording device are relatively close. This highlights the influence of sessions and devices on the learning process, likely due to environmental noise, device setup, and physiological or behavioral variations over time. Finally, the analysis of the three devices used for data collection indicates that cross-device authentication is feasible when the feature extractor is trained on subjects who have collected data with both devices.

    \item \textbf{Enhancing Performance (Sec \ref{sec:performance}):} We investigated the trade-off between verification time and performance. For example, by increasing the verification time from 1 to 4 seconds, the error rate decreased by approximately 40\%. Additionally, we examined the trade-off between enrollment sessions and error rates, observing that after the second enrollment session, the improvement in error rate diminished.

    \item \textbf{Channel Reduction (Sec \ref{sec:consumer}):} To evaluate the system's performance under conditions resembling real-world scenarios, we reduced the number of channels in the PEERS dataset---collected using medical-grade devices---from 93 to 14, 7, and 4. These reductions correspond to sensor placements commonly found in consumer-grade EEG devices. The results showed a relatively small increase in error compared to the notable reduction in channels.

    \item \textbf{Industrial Standard Comparison (Sec \ref{sec:comparative}):} For the first time, we compare brainwave authentication performance against international biometric standards. At required security levels, the false rejection rate (FRR) exceeds acceptable thresholds. However, our results show a logarithmic relationship between error rate and training subjects, suggesting that with more training data, these standards may be met at a practical FRR.

    \item \textbf{Open Sourced Code:} We published our source code, which, along with the public dataset, makes results easily reproducible and facilitates future error rate reductions by other researchers~\footnote{\scriptsize{https://github.com/kit-ps/NeuroShield/}}.

\end{itemize}

\section{Background and Related Work}
EEG measures electrical activity produced by neurons firing in the human brain, typically recorded at the scalp using specialized sensors \cite{niedermeyer2005electroencephalography, hu2019eeg}. German physiologist and psychiatrist Hans Berger recorded the first human EEG in 1924 for medical purposes, which remains one of its primary applications \cite{gupta2022review, adeli2010automated, haas2003hans}. Beyond medical use, EEG technology has advanced to support Brain-Computer Interfaces (BCIs), 
systems that enable direct communication between the brain and external devices \cite{kawala2021summary}. Recently, EEG data has also been found to include unique features linked to user identity, making it a promising method for user authentication \cite{gui2019survey}. Brainwave based authentication is hands-free, resistant to shoulder surfing, revocable, renewable, and supports continuous authentication.

Biometric authentication modalities should meet certain requirements to enable widespread usage. These requirements include universality (the modality must be present in all individuals), distinctiveness (it must uniquely identify an individual), permanence (it should remain consistent over time), and collectability (it must be measurable with available technology). Additional factors include performance (accuracy, speed, and resource efficiency), acceptability (user willingness to utilize the system), and circumvention resistance (difficulty in forging or bypassing)~\cite{rui2018survey}. 

EEG is universal because each person’s brain produces EEG signals~\cite{marcel2007person}. These signals exhibit distinctiveness due to individual brainwave patterns, making them uniquely identifiable~\cite{kaliraman2019use}. Advances in technology are rapidly improving collectability, with new EEG headsets entering the market annually, offering better signal quality and usability~\cite{yang2017usability}. However, without real-time implementation, it is difficult to comprehensively assess acceptability and circumvention resistance, though some research attempts have been made in this direction~\cite{fallahi2024usability,wang2022cancellable}.

Currently, the primary focus of research is on performance and permanence, as these are challenging aspects of EEG-based authentication. The two are closely interrelated, with performance metrics reflecting the system's ability to maintain consistent accuracy over time. Factors such as hair growth, slight variations in sensor placement on the scalp, brain states (e.g., stress), and environmental conditions (e.g., noise) can significantly impact both performance and permanence in EEG-based authentication.
However, most research studies evaluate the performance of brainwave authentication using single-session datasets (cf. \cite{schons2018convolutional, bidgoly2022towards, arias2023performance,fallahi2023brainnet}), which neglect session invariance (permanence) completely. 
This limitation raises concerns that models may overfit to session-specific characteristics rather than capturing subject-unique features. 
Additionally, some studies (cf. \cite{debie2021session,chaurasia2024neuroidbench}) consider short time intervals, limited to periods of a few days only, which may only partially account for the aforementioned issues. 
While a limited number of papers explore time intervals spanning more than a week, these studies (cf. \cite{das2016eeg,wu2018eeg,maiorana2017longitudinal,maiorana2021learning}) rely on non-public datasets and omit sharing their source code, hindering reproducibility and further research. 
Moreover, such studies have been conducted with a relatively small sample size and limited sessions (up to 54 subjects~\cite{debie2021session} and 270 sessions~\cite{maiorana2017longitudinal}), which increases the risk of overfitting and restricts the generalizability of the results.

Therefore, a more comprehensive investigation of performance and permanence is needed using larger multi-session datasets to avoid overfitting. This would enable the examination of key parameters affecting performance over time and provide insights into improving brainwave authentication. Additionally, identifying the main challenges and ensuring reproducibility through open-source code sharing are essential for advancing the field.
\section{Method}
\label{sec:method}
Biometric systems, including brainwave authentication, involve four key stages: data collection, preprocessing, feature extraction, and feature comparison. 
Like other biometric systems, they operate in two phases: enrollment and verification.
In the enrollment phase, brainwave signals are collected from the user to generate a reference profile, known as a template. During verification, performed after enrollment, brainwave signals are captured again and compared with the reference profile to confirm the user's identity.

\begin{table}
\caption{Demographic and session distribution of the dataset.}
\label{tab:dataset_summary}
\begin{tabular}{ll}
\toprule
Metric & Value \\
\midrule
Unique subjects & 345 \\
Unique sessions & 6007 \\
Average sessions per subject (± SD) & 17.41 $\pm$ 10.48 \\
Female & 161 (46.7\%) \\
Male & 136 (39.4\%) \\
Unspecified sex & 48 (13.9\%) \\
Age, years (mean ± SD) & 27.0 $\pm$ 15.9 \\
Age, years (range) & 17.0--85.0 \\
\bottomrule
\end{tabular}
\end{table}

\subsection{Dataset Description}
Our study utilizes the Penn Electrophysiology of Encoding and Retrieval Study (PEERS) dataset~\cite{kahana2024penn}, provided by the University of Pennsylvania. Collected between 2010 and 2020, the dataset focuses on exploring EEG correlates of memory processes, particularly during tasks involving word memorization. In these tasks, participants are shown a sequence of words and asked to memorize them as stimulus. Stimulus presentation was synchronized with EEG data collection using event markers. We used the timestamps of word presentations on the screen to segment the EEG data into individual samples. EEG data, recorded as time series of scalp voltage signals, are collected during the sessions at sampling rate of 500 Hz. The dataset includes approximately 8.7~TB of EEG data from 345 subjects, each with more than one session, totaling over 6007 sessions. The participants range in age from 17 to 85 years, with an average age of 26.98 $\pm$ 15.86 years. The dataset includes 161 female and 136 male participants, while gender information for 48 subjects is unavailable (cmp. Table ~\ref{tab:dataset_summary}). The PEERS dataset is particularly suited for our research as it includes multiple sessions per subject, enabling the investigation of inter-session variability, a critical factor for ensuring the robustness of real-world biometric systems. Data collection was conducted using three distinct medical-grade headsets: the 129-channel Geodesic Sensor Net (GSN 200 model)\footnotemark[2] , the 129-channel HydroCel Geodesic Sensor Net.\footnotemark[2], and the 128-channel BioSemi
\footnote{\scriptsize\url{https://www.biosemi.com/headcap.htm}} headcap using the Biosemi ActiveTwo acquisition system. These diverse devices further enhance the dataset's relevance by reflecting real-world variability in EEG acquisition systems.

\footnotetext[2]{\scriptsize\url{https://med.stanford.edu/content/dam/sm/lucasmri/documents/16_0824_EGI_geodesic_sensor_net.pdf}}

\subsection{Preprocessing}
As the channel layout of the different headsets varied, we unified them by mapping the channels to a well-known extended version\footnote{\scriptsize\url{https://mne.tools/stable/auto_tutorials/intro/40_sensor_locations.html}} of the ten-twenty electrode system~\cite{silverman1963rationale}, which includes 93 channels. Subsequently, the EEG data channels were reordered for each recording to align with the expected order during evaluation across all three devices. 
Following this, and in line with best practices for preprocessing in brainwave authentication~\cite{chaurasia2024neuroidbench,arias2023performance,peng2019eeg,maiorana2021learning}, a common average reference (CAR) was applied to reduce spatially correlated noise and improve signal quality. Next, we applied a bandpass filter between 1.0 Hz and 50.0 Hz using a zero-phase finite impulse response (FIR) filter with default parameters, as implemented in MNE-Python~\cite{gramfort2014mne}, to focus on frequency bands that contain EEG data relevant for authentication~\cite{suppiah2018biometric}. As the 50 Hz power-line frequency lies close to the upper end of the preserved frequency band, we additionally applied a notch filter at 50 Hz to suppress potential residual line noise~\footnote{Residual power-line interference in EEG recordings typically occurs at 50 Hz (e.g., most of Europe) or 60 Hz (e.g., North America). We bandpass-filtered the data between 1 and 50 Hz and then applied notch filters at 50 Hz and 60 Hz to suppress residual line noise at both mains frequencies; because 60 Hz lies outside the preserved band, its notch filter effect on the data is negligible.}. After filtering, the data was normalized using a robust normalization approach, where the median was subtracted and the result divided by the interquartile range (IQR), mitigating the influence of outliers while preserving the temporal structure of the signal. The data then consisted of 1-second non-overlapping samples aligned to stimulus timestamps. Each sample was stored together with metadata including session date and hardware type.

\subsection{Feature Extraction}
After preprocessing, each sample contains 46,500 values (93 channels × 500 Hz). Due to the high dimensionality and variability of EEG data, direct comparison is not feasible. Therefore, feature extraction is necessary to capture subject-specific information across sessions. Two main approaches exist: the first is handcrafted feature extraction, which uses predefined formulas based on expert knowledge and does not require training data. The second approach uses deep learning techniques, which require a separate training set to learn feature representations. In this method, the feature extractor is trained on labeled data to produce representations that generalize to new subjects.

\textbf{handcrafted Feature Extraction:} We use Power Spectral Density (PSD) to transform EEG signals from the time to frequency domain, capturing patterns linked to brain dynamics. PSD is commonly used in brainwave-based authentication systems~\cite{khalil2024unlocking,arias2023performance,chaurasia2024neuroidbench}, applying Fourier transformation to quantify signal power across frequency components~\cite{welch1967}. We also extract AutoRegressive (AR) model coefficients, which capture temporal dependencies by modeling each point as a linear combination of previous values. AR features provide compact representations for authentication \cite{arias2023performance,nai2004classification}. Finally, we combine PSD and AR features to incorporate both frequency and temporal information.

\textbf{Deep learning feature extraction:} We applied metric learning, which trains a network to map inputs into a feature space where samples with the same label are close and those with different labels are far~\cite{kulis2013metric}. This method uses an encoder to generate embeddings and a loss function that enforces distances based on labels. Metric learning losses are distance-based or softmax-based. Distance-based losses, operating on pairs, triplets, or mined batches, optimize embedding distances directly~\cite{horiguchi2019significance}. Softmax-based losses encourage class separation by clustering embeddings and generally need larger datasets to generalize well~\cite{horiguchi2019significance}.

For softmax-based approaches, we selected two loss functions: (1) SoftTripleLoss, which addresses intra-class variations by introducing multiple learnable centers per class; and (2) ArcFaceLoss, which applies angular margin constraints to improve inter-class separability and robustness to noise. For distance-based approaches, we selected three loss functions: (1) Triplet Loss \cite{schroff2015facenet}, which minimizes the distance between an anchor and a positive sample while maximizing the distance from a negative sample; (2) LiftedStructureLoss \cite{oh2016deep}, which uses all positive and negative pairs within a batch and penalizes violations of distance constraints; and (3) SupConLoss (Supervised Contrastive Loss)~\cite{khosla2020supervised}, which extends contrastive learning by using class labels to group embeddings of the same class together and separate embeddings of different classes. This supervision helps SupConLoss align embeddings more effectively across the batch.

For Triplet Loss, we adopt semi-hard online triplet mining \cite{schroff2015facenet}, which is recommended to emphasize informative (non-trivial) triplets and improve convergence. LiftedStructureLoss and SupConLoss are pair-based (and not triplet-based); accordingly, we employ Multi-Similarity mining \cite{wang2019multi} to select hard positive and negative pairs based on cosine similarity. ArcFaceLoss and SoftTripleLoss are proxy-based margin losses (softmax-based) that operate on class centers and therefore do not require mining, as they do not consume explicit pair or triplet indices. Samples are drawn across all sessions in the training set, and positive pairs include both within-session and cross-session pairs of the same subject. This allows the model to minimize both intra-session and inter-session embedding distances, supporting continuous authentication (stability within sessions) as well as multi-session authentication (robustness across sessions).

%During distance-based training, samples are matched by subject label across all sessions, so positive pairs may originate from different sessions of the same subject.

For each loss function, different encoder architectures were explored to optimize feature extraction. Specifically, we implemented six backbone EEG models — ResNet1D~\cite{zheng2022task}, ShallowNet~\cite{schirrmeister2017deep}, DeepConvNet~\cite{schirrmeister2017deep}, EEGNet~\cite{lawhern2018eegnet}, LSTM~\cite{zhang2021deep}, and GRU~\cite{zhang2021deep} — as encoder networks. These models were selected to cover a diverse range of architectures, including convolutional and recurrent networks, and to balance model complexity and representational capacity for EEG signal processing. Details of the neural network architecture are in Appendix~\ref{A:B}.

\subsection{Feature Comparison}
After reducing brainwave samples to lower dimensional embeddings through feature extraction, the extracted features should be compared, and a similarity score calculated to evaluate verification attempts. Two main approaches can be used. The first approach applies distance metrics; we select Euclidean, Cosine, and Manhattan distances because Euclidean captures absolute differences, Cosine measures angular similarity independent of magnitude, and Manhattan is more robust to outliers~\cite{fallahi2024beyond, jiang2023cancelable}. The second approach trains binary and one-class classifiers to distinguish between positive and negative samples for each subject. However, binary classifiers require negative samples, which should not come from subjects included in the evaluation. We employ several binary classifiers, including logistic regression, linear discriminant analysis (LDA), support vector machine (SVM) with linear kernel, stochastic gradient descent (SGD) classifier with radial basis function (RBF) kernel approximation, and random forests. One-class SVM is also tested, although it is less suitable when negative classes are available; it is included due to its use in related work~\cite{maiorana2021learning}. As shown in face recognition, distance-based approaches are expected to perform better if sufficient data is available to train the feature extractor~\cite{schroff2015facenet}, since metric learning aims to minimize distances between similar samples. The most suitable comparison method will be selected via systematic hyperparameter tuning, as described in the Section~\ref{sec:benchmarke}.

\subsection{Threat model}
Since EEG signals are not externally observable, shoulder surfing is not possible, and the adversary cannot impersonate a victim. Accordingly, authentication studies typically assume that adversaries attempt to gain access using their own brainwave recordings~\cite{bidgoly2022towards,arias2023performance}. Meanwhile, our goal is to investigate the reliability of brainwave-based authentication over time (multi-session setting), and to identify potential bottlenecks affecting this reliability. The low reliability of brainwave signals across sessions contributes to a high false acceptance rate, which poses a significant security threat and can be assessed using zero-effort attacks \cite{mansfield2002best}. A zero-effort attack in biometric authentication occurs when an adversary uses their own biometric sample in an attempt to gain unauthorized access to another individual's account. In this context, the security objective of the biometric system is to minimize the probability that the authentication algorithm incorrectly verifies the adversary’s identity as the legitimate user. Specifically, the likelihood of successful verification of a falsely claimed identity, based on the adversary's sensor readings, should be negligible. Therefore, zero-effort attacks are commonly used in studies on multi-session reliability involving brainwave data~\cite{maiorana2021learning,maiorana2017longitudinal,das2016eeg,wu2018eeg,chaurasia2024neuroidbench,debie2021session}.

\subsection{Performance Metrics}
We used the Equal Error Rate (EER) as a summary metric, indicating the point where the False Acceptance Rate (FAR) and the False Rejection Rate (FRR) are equal. The FAR represents the proportion of unauthorized attempts incorrectly accepted by the system, while the FRR refers to the proportion of legitimate attempts incorrectly rejected. EER has been widely used in biometric authentication papers~\cite{rui2018survey,ayeswarya2024comprehensive} to report results and ensure comparability. However, in practical usage, it is preferred to keep FAR at very low rates to ensure system security, even at the cost of a higher FRR. Therefore, we also report the FRR at a FAR threshold of 1\% for comparison with other brainwave authentication studies~\cite{arias2023performance,fallahi2023brainnet,beyondreality2022}, as well as at 0.1\%\footnote{\scriptsize\url{https://pages.nist.gov/800-63-3/sp800-63b.html}} and 0.01\%\footnote{\scriptsize\url{https://pages.nist.gov/800-63-4/sp800-63b.html}} to align with biometric standards.

\begin{table}[ht]
  \centering
  \caption{Results of hyperparameter trials on the validation set (15 subjects): number of pipeline occurrences within the top 1,000 trials, average EER computed for pipelines appearing in the top 1,000, and best EER across all 10,000 trials.}
  \label{tab:performance-category}
  \begin{tabular}{llccc}
    \toprule
    & Pipeline & Count & Avg & Best \\
    \midrule
    \multirow{31}{*}{\rotatebox{90}{Feature Extraction}}
        & ResNet1D + SupConLoss & 98 & 10.20 & 8.05 \\
        & ResNet1D + ArcFaceLoss & 65 & 10.65 & 8.09 \\
        & ResNet1D + LiftedStructureLoss & 107 & 10.78 & 8.39 \\
        & ShallowNet + SoftTripleLoss & 77 & 11.13 & 8.97 \\
        & ResNet1D + TripletMarginLoss & 99 & 11.36 & 9.20 \\
        & ShallowNet + LiftedStructureLoss & 34 & 10.85 & 9.25 \\
        & ShallowNet + SupConLoss & 77 & 11.50 & 9.38 \\
        & ResNet1D + SoftTripleLoss & 54 & 11.79 & 9.66 \\
        & ShallowNet + ArcFaceLoss & 72 & 12.12 & 9.69 \\
        & DeepConvNet + SoftTripleLoss & 90 & 12.19 & 10.23 \\
        & GRU + SoftTripleLoss & 36 & 11.43 & 10.35 \\
        & LSTM + SoftTripleLoss & 28 & 11.87 & 10.72 \\
        & DeepConvNet + ArcFaceLoss & 26 & 12.10 & 11.11 \\
        & ShallowNet + TripletMarginLoss & 47 & 12.30 & 11.50 \\
        & GRU + SupConLoss & 28 & 12.62 & 12.04 \\
        & DeepConvNet + SupConLoss & 10 & 12.69 & 12.34 \\
        & LSTM + LiftedStructureLoss & 9 & 12.71 & 12.59 \\
        & DeepConvNet + LiftedStructureLoss & 7 & 12.68 & 12.63 \\
        & GRU + LiftedStructureLoss & 8 & 12.69 & 12.65 \\
        & EEGNet + LiftedStructureLoss & 9 & 12.93 & 12.75 \\
        & GRU + ArcFaceLoss & 19 & 12.89 & 12.79 \\
        & EEGNet + ArcFaceLoss & 0 & -- & 13.18 \\
        & LSTM + ArcFaceLoss & 0 & -- & 13.21 \\
        & EEGNet + SupConLoss & 0 & -- & 13.36 \\
        & LSTM + SupConLoss & 0 & -- & 13.61 \\
        & DeepConvNet + TripletMarginLoss & 0 & -- & 13.88 \\
        & EEGNet + SoftTripleLoss & 0 & -- & 14.49 \\
        & EEGNet + TripletMarginLoss & 0 & -- & 15.00 \\
        & LSTM + TripletMarginLoss & 0 & -- & 15.55 \\
        & GRU + TripletMarginLoss & 0 & -- & 18.74 \\
        & PSD + AR & 0 & -- & 19.23 \\
        & AR & 0 & -- & 19.43 \\
        & PSD & 0 & -- & 29.34 \\
    \addlinespace
    \midrule
    \multirow{9}{*}{\rotatebox{90}{Feature Comparison}} 
               & Euclidean Distance             &  86  & 11.18 & 8.05 \\
               & Cosine Distance                & 168  & 10.75 & 8.05 \\
               & Manhattan Distance             & 105  & 10.80 & 8.17 \\
               & Logistic Regression            & 171  & 11.70 & 8.36 \\
               & LDA                             & 222  & 11.67 & 8.56 \\
               & Linear SVM                      & 106  & 11.95 & 8.86 \\
               & SGD RBF Kernel                  & 80   & 11.74 & 9.58 \\
               & Random Forest                   & 62   & 12.45 & 11.49 \\
               & One-Class SVM                   & 0    & --  & 21.68 \\
    \bottomrule
  \end{tabular}
\end{table}

\section{Results and Discussion}
In order to evaluate brainwave authentication performance over time, we first select a pipeline comprising a feature extractor and a comparison metric through hyperparameter tuning. We then investigate important parameters such as test size, time interval, recording hardware, and subject embeddings visualized as distinct clusters through t-SNE plots. Next, to move toward consumer-grade devices, we analyze performance with a reduced number of channels and  contextualize our findings with state-of-the-art methods and authentication standards to identify potential directions for future research.

\subsection{Benchmarking Feature Extraction and Compression}
\label{sec:benchmarke}
After introducing  dataset and preprocessing methods we now want to turn to selecting suitable feature extraction and compression techniques, that are essential to complete the authentication pipeline.

%\subsubsection{Experiment setup} Our claim is that the proposed method supports both inter-subject and inter-session generalization. Therefore, this is reflected by the data split and evaluation protocol. We divided the 345 available subjects into non-overlapping sets: 230 subjects for training the deep learning-based feature extractor, 15 for validation, and 100 randomly selected subjects for evaluation, on whom system performance is measured. In all experiments, unless specified otherwise, the first session of each subject was used for enrollment, and the remaining sessions of the same subject for verification. This setup ensures inter-subject evaluation and shows that the model does not require retraining when new users are added. Additionally, separating enrollment and verification by session enables the assessment of inter-session generalization.

\subsubsection{Experiment setup} We hypothesize that the proposed method supports both inter-subject and inter-session generalization. Inter-subject generalization refers to the ability of a model trained on a subset of users to perform well on previously unseen users. Inter-session generalization refers to the ability of the model to embed data from different sessions of the same user close to each other in the feature space. Therefore, this is reflected by the data split and evaluation protocol. We divided the 345 available subjects into non-overlapping sets: 230 subjects for training the deep learning-based feature extractor, 15 for validation, and 100 randomly selected subjects for evaluation, on whom system performance is measured. In all experiments, unless specified otherwise, the first session of each subject in the evaluation set was used for enrollment, and the remaining sessions of the same subject for verification. This split ensures inter-subject evaluation, since no subject appears in more than one of the training, validation, and evaluation sets, and shows that the model does not require retraining when new users are added. Additionally, separating enrollment and verification by session enables the assessment of inter-session generalization.

\subsubsection{Hyperparameter tuning} From each combination of model architectures and loss functions (6\texttimes5) mentioned in the methods (Section \ref{sec:method}), we trained a model using hyperparameter tuning on the optimizer, learning rate, batch size, and embedding dimension based on the training set, guided by performance on the validation set. The entire process took approximately 144 hours on four NVIDIA GeForce RTX 4090 GPUs.

After obtaining the feature extraction models, we conducted general hyperparameter tuning to select the best combination of feature extraction and feature comparison methods. The tuning process was guided by the validation set and employed a random search over 10,000 trials. To provide a clearer overview, Table~\ref{tab:performance-category} reports the results for the top 10\% of trials on the validation set, thereby reducing the influence of high-error outliers on the average results. The hyperparameter tuning identified the ``ResNet1D + SupConLoss + Euclidean distance" pipeline as the best configuration.
This pipeline was then evaluated on a test set of 100 subjects, resulting in an EER of 10.28\% ± 6.3\%. \textit{For the remainder of this paper, we use this pipeline as the default authentication pipeline.}

\subsubsection{Feature Extraction Methods} Table~\ref{tab:performance-category} shows that our dataset is sufficient for deep learning approaches to outperform handcrafted features, which result in significantly higher error. Among the encoders, ResNet1D consistently achieved the top three ranks, followed by ShallowNet, which outperformed the remaining models. This suggests that convolutional networks generally perform better than recurrent architectures. Among the convolutional models, simpler architectures like ShallowNet may benefit from reduced overfitting, while ResNet1D leverages residual connections to achieve better performance than models such as DeepConvNet. Although EEGNet is relatively shallow, its specialized filters for frequency and spatial information did not translate into strong performance in our evaluation. Regarding loss functions, although SoftTripleLoss---with its multiple learnable centers per class---performed best with shallower and recurrent backbones (ShallowNet, DeepConvNet, LSTM, GRU), ResNet1D combined with SupConLoss achieved the highest overall performance across all architectures. SupConLoss combines the strengths of distance-based and softmax-based approaches by leveraging all positive and negative pairs in the batch, guided by class labels. This batch-level supervision enforces strong intra-class cohesion and inter-class separation, which ResNet1D---with its higher capacity---can fully exploit to learn highly discriminative EEG embeddings

\subsubsection{Comparison Methods} The results (Table~\ref{tab:performance-category}) suggest that, in general, distance-based functions outperform shallow classifier approaches for feature comparison. This is not surprising, as loss functions in metric learning are designed to minimize distances between embeddings, typically using cosine or Euclidean distance. In contrast, shallow classifiers either require negative samples for training or, in the case of one-class classification, often show relatively poor performance due to the instability associated with one-class models. This is also why state-of-the-art face recognition methods use distance functions for feature comparison.

\begin{figure*}
    \centering
    \includegraphics[width=0.9\linewidth]{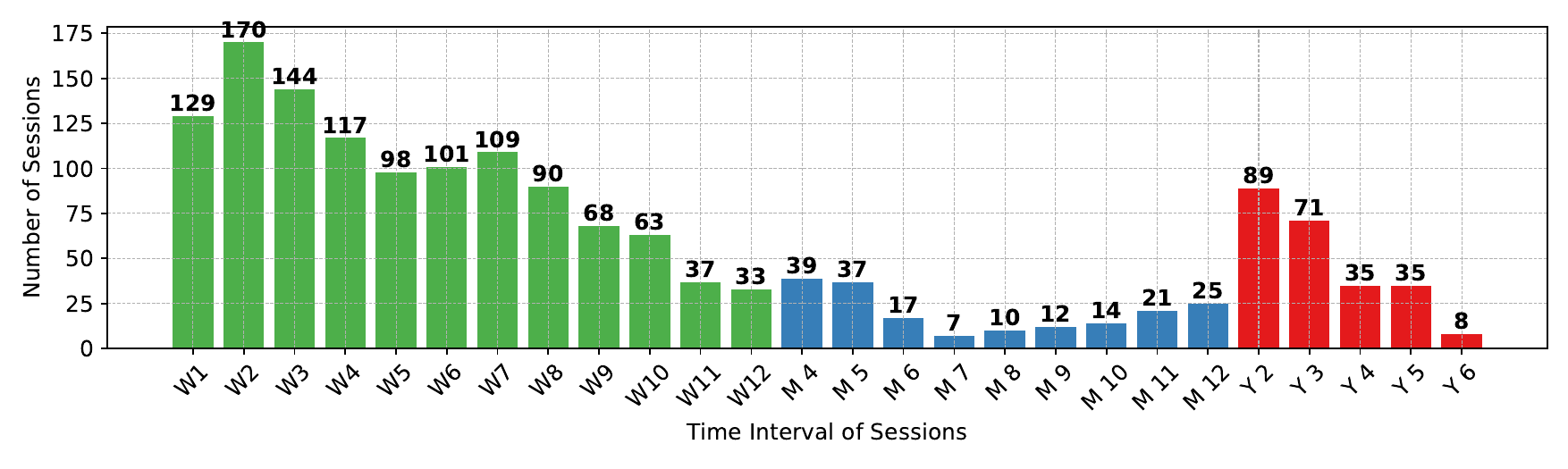}
    \caption{Number of sessions in the evaluation dataset  post-enrollment. \(W\), \(M\), and \(Y\) denote week, month, and year. %respectively. 
    Labels (e.g., \(W3\)) mark the number of sessions in that time unit, starting the day after the previous unit ends (e.g., \(W3\): days 15–21).}
    \label{fig:a1}
\end{figure*}

\begin{figure}
    \centering
    \includegraphics[width=0.8\linewidth]{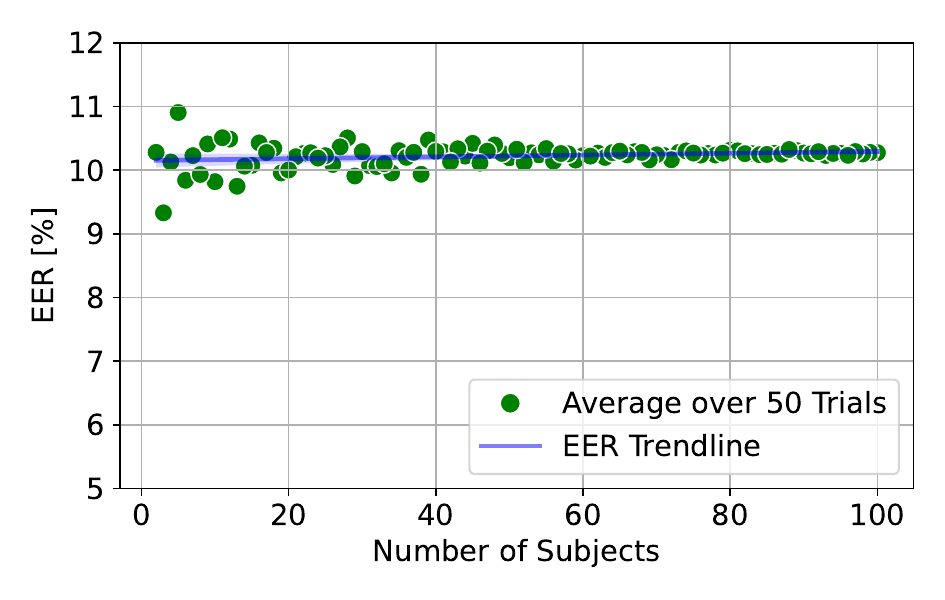}
    \caption{EER vs. Number of Subjects: Each blue dot represents the average of 50 instances of randomly selecting \(N\) subjects and calculating the EER. The dot size represents the standard deviation of the EER values across these 50 calculations.}
    \label{fig:subject_number1}
\end{figure}

\subsection{Detailed Performance Analysis}
\label{sec:analysis}
With the best-performing pipeline identified, the EER remains significantly lower than the 50\% chance level but still reaches 10.28\%, which is higher than the performance required for real-world deployment. This result may indicate that brainwaves are not sufficiently unique for authentication or that other underlying factors need to be understood to develop effective solutions. Therefore, we aim to investigate which parameters influence multi-session authentication more deeply by addressing the following research questions:

\begin{enumerate}
    \item How does altering the number of participants in an EEG-based authentication evaluation set affect the overall comparability of the evaluation?  To investigate this, we analyze how varying the number of evaluation subjects from 2 to 100 affects the EER.
    
    \item How does EEG-based authentication performance change over time, and what does this suggest about brainwave pattern stability? To explore this, we examine how EER is influenced by varying verification and enrollment intervals.
    
    \item How effectively are subjects clustered based on their embedding space derived from the feature extractor? This explores the visualization of embedding space.

    \item How does different recording hardware affect the performance of brainwave-based authentication, and what are the best practices for ensuring reliability? This question is investigated by measuring the EER on hardware-specific and generalized models as well as evaluating cross-hardware authentication scenario.
    
\end{enumerate}

\begin{figure*}
    \centering
    \includegraphics[width=1\linewidth]{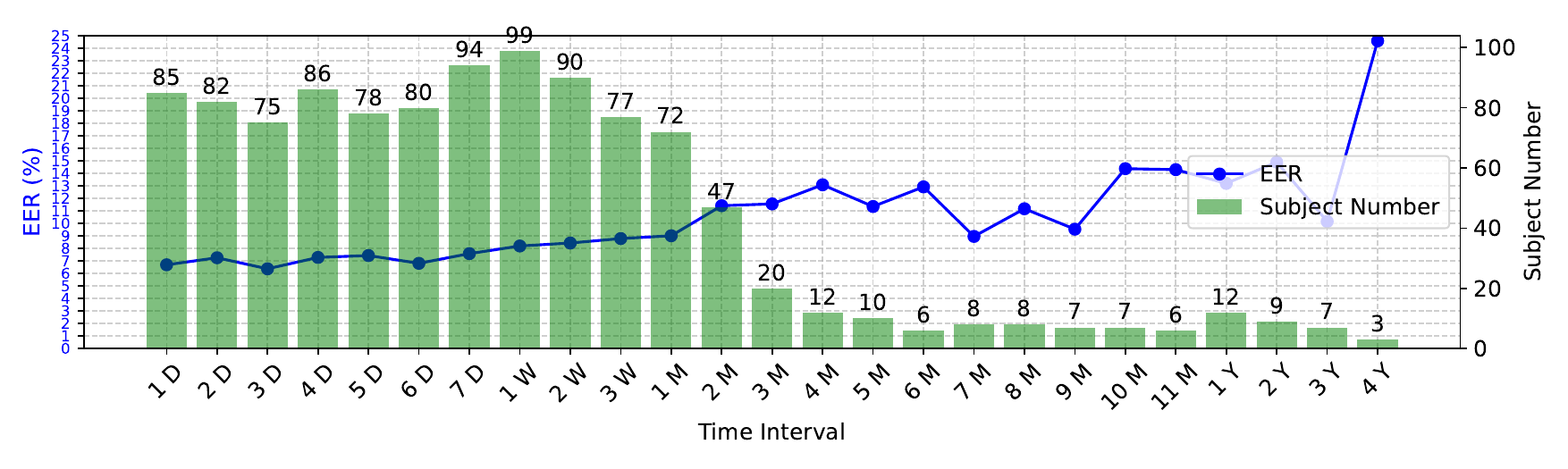}
    \caption{Relationship between the time interval between sessions and EER. \(D\) represents a day, \(W\) a week, \(M\) a month, and \(Y\) a year. The value after each symbol indicates the start time interval (except for days). For example, \(W1\) represents sessions with a time interval of 8–14 days after the first session.}
    \label{fig:a2}
\end{figure*}

\subsubsection{Effect of the Number of Evaluation Subjects on EER} 
When comparing the results of different experiments using metrics such as EER,  number of subjects in the evaluation is typically reported as metadata.  A larger test set generates more impostor scores in zero-effort scenarios, making it valuable to investigate the relationship between the number of subjects in the evaluation set and the resulting error. This analysis provides insights for better comparisons of results across different studies or our follow-up experiments with smaller test sets, such as when evaluating performance using headset-specific recordings. To achieve this, we randomly select \(N\) subjects, where \(N\) ranges from 2 to 99 (starting at \(N=2\) since at least one genuine user and one impostor are required). For each value of \(N\), we compute the EER, repeat the procedure 50 times, and report the average EER per \(N\).

Figure \ref{fig:subject_number1} presents the results of the experiment. The analysis revealed no significant relationship between the number of subjects and the EER. These findings align with observations in face recognition systems. For example, Friedman et al. \cite{friedman2022biometric} investigated how test size affects identification and authentication tasks. They reported a linear decline in identification performance with the logarithm of enrollment size but found that the EER for authentication tasks remained stable. This stability occurs because the EER is based on the Receiver Operating Characteristic (ROC) curve, which balances the FAR and FRR derived from impostor and genuine score distributions. While larger subject pools better define these distributions, they do not significantly influence the ROC. Therefore, we can conclude that having a higher number of subjects in the test set does not necessarily make authentication more challenging. However, both scenarios indicate that test sets with a small number of participants exhibit higher deviations in EER, which can lead to unreliable or misleading results. This underscores the importance of using larger test sets to obtain more dependable evaluations.

\begin{figure}
    \centering
    \includegraphics[width=0.8\linewidth]{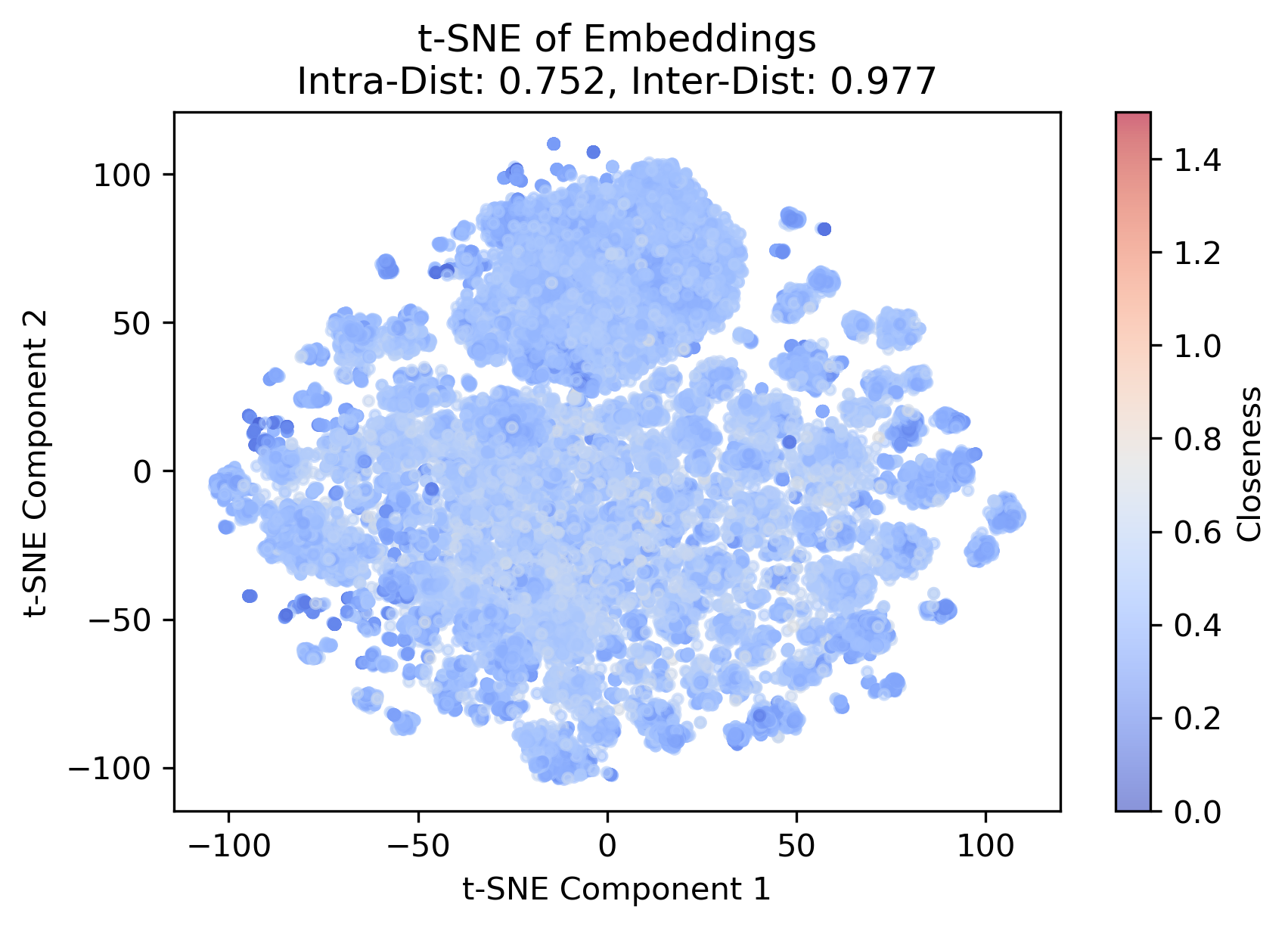}
    \caption{t-SNE visualization of embeddings. Color represents the average cosine distance to the top 5 nearest samples from the same subject but different session. Warm colors indicate greater distances. Considering all samples, the average intra-class distance is 0.75 and the average inter-class distance, 0.97.
    }
    \label{fig:SNE1}
\end{figure}

\subsubsection{Correlation Between EER and Time Interval}
To investigate permanence, we first visualize the distribution of evaluation session intervals in Figure \ref{fig:a2}. The test data includes a total of 1,699 unique sessions available for 100 evaluation subjects, where the first session (or the oldest session) per subject is considered as the enrollment session (100 sessions). The results show that the highest concentration of sessions occurred within the first week, with 129 sessions (8\%) recorded. By the end of the first month (weeks 0 to 3), this number increased to 560 sessions (35\%), indicating that one-third of all sessions were conducted in the first month. The second month (weeks 4 to 7) accounted for 398 sessions (24.8\%), while the third month (weeks 8 to 11) added 201 sessions (12.5\%). Beyond the third month, session frequency declined further, with 182 sessions (11.3\%) recorded from Month 4 to the end of the first year. In the second year, session numbers were 89 sessions (5.5\%), and remained similar at 71 sessions (4.4\%) in the third year. The remaining years combined accounted for 4.8\%.

To investigate the correlation between time intervals between sessions and authentication error, we extracted all possible pairs of sessions within specific time intervals and calculated the EER for each. Figure \ref{fig:a2} shows the results, demonstrating a positive correlation between time interval and error rate, as expected. Verifying identity becomes harder over time. The EER after one day is 6.7\%, increasing to 7.6\% after seven days. However, within the second month, the EER rises to 7.2\%, which remains lower than the previously reported average of 10.28\%, indicating stable performance over a reasonable time interval. Beyond the second month, the EER gradually increases, reaching 14.3\% after one year. Performance remains relatively stable in the second and third years. In the fourth and fifth years, performance deteriorates significantly, with EER values of 24.6\% and 47.5\%, respectively. However, due to the limited data for these time points, the results are not fully reliable. Nonetheless, even the doubling of EER after just one year suggests substantial degradation, indicating the need to update the enrollment set at shorter time intervals.

\begin{figure}
    \centering
    \includegraphics[width=0.8\linewidth]{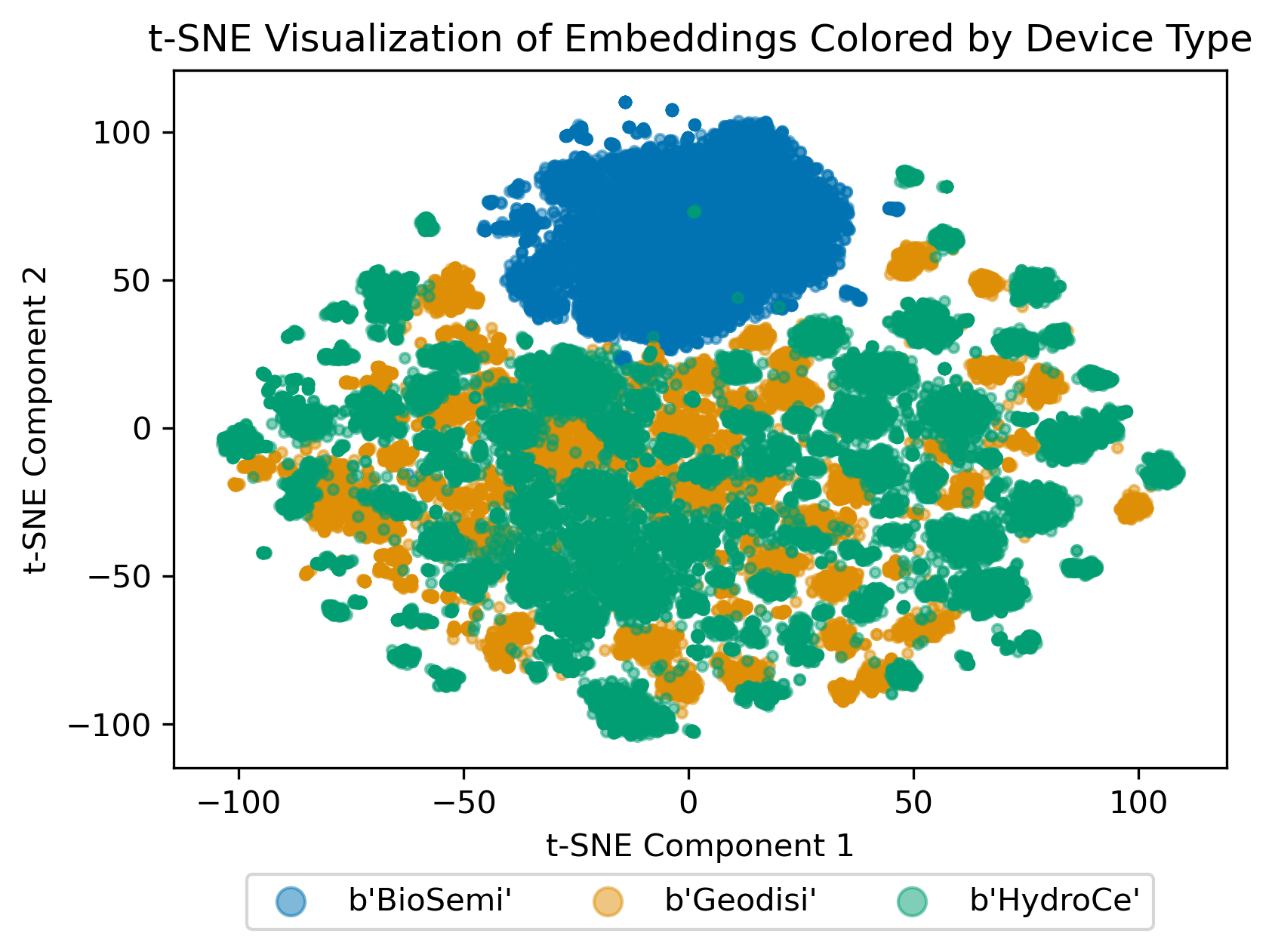}
    \caption{t-SNE visualization of embeddings.  Colors indicate different EEG headsets used for data collection.}
    \label{fig:SNE2}
\end{figure}

\subsubsection{Visualize Embedding Space}
To better understand how well users are represented in distinct clusters, we project the embeddings into a 2D space using t-Distributed Stochastic Neighbor Embedding (t-SNE) and visualize the results. Figure~\ref{fig:SNE1} shows that the metric learning approach successfully separates subjects: the average distance between a subject’s own samples is significantly lower (0.75) than the average distance to other subjects’ samples (0.97), indicating the model’s ability to capture individual differences. However, we observe two separate colonies within the embedding space, leading us to hypothesize that these colonies could be due to differences in the hardware used during data capture. This hypothesis is confirmed by Figure~\ref{fig:SNE2}, which reveals that the data captured using BioSemi hardware is located separately from the rest, further illustrating the significant influence of hardware on the embedding distribution. The effect of hardware on the results will be discussed further in the next section.

\begin{figure}
    \centering
    \includegraphics[width=1\linewidth]{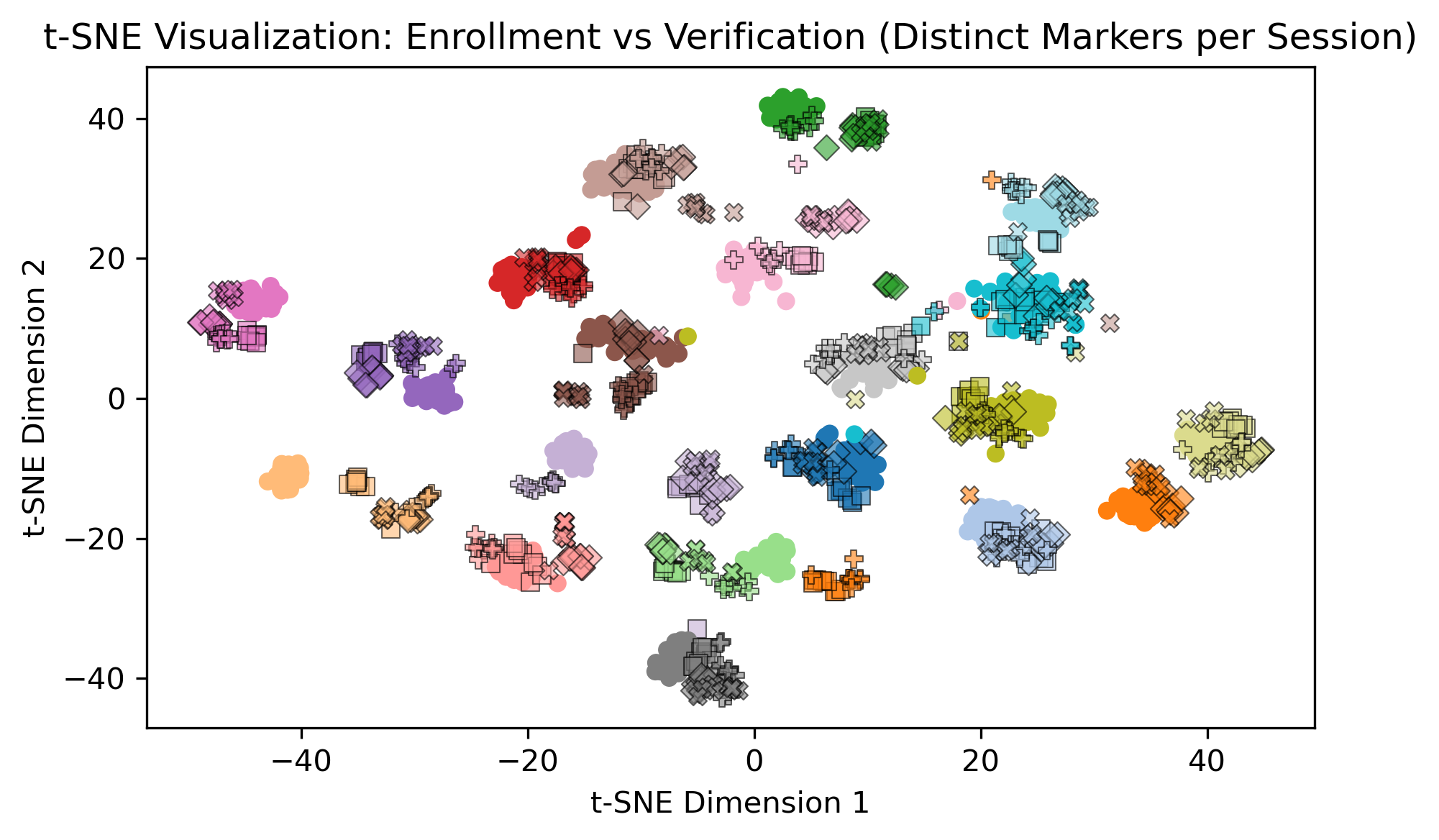}
    \caption{t-SNE visualization of embeddings for 20 randomly selected subjects, with each subject assigned a unique color for a clear distinction. Enrollment samples are depicted as circles without borders, while verification samples are represented by various shapes with black borders. Also, each unique shape corresponds to a distinct session for verification samples.}
    \label{fig:highl1}
\end{figure}
\begin{figure}
    \centering
    \includegraphics[width=1\linewidth]{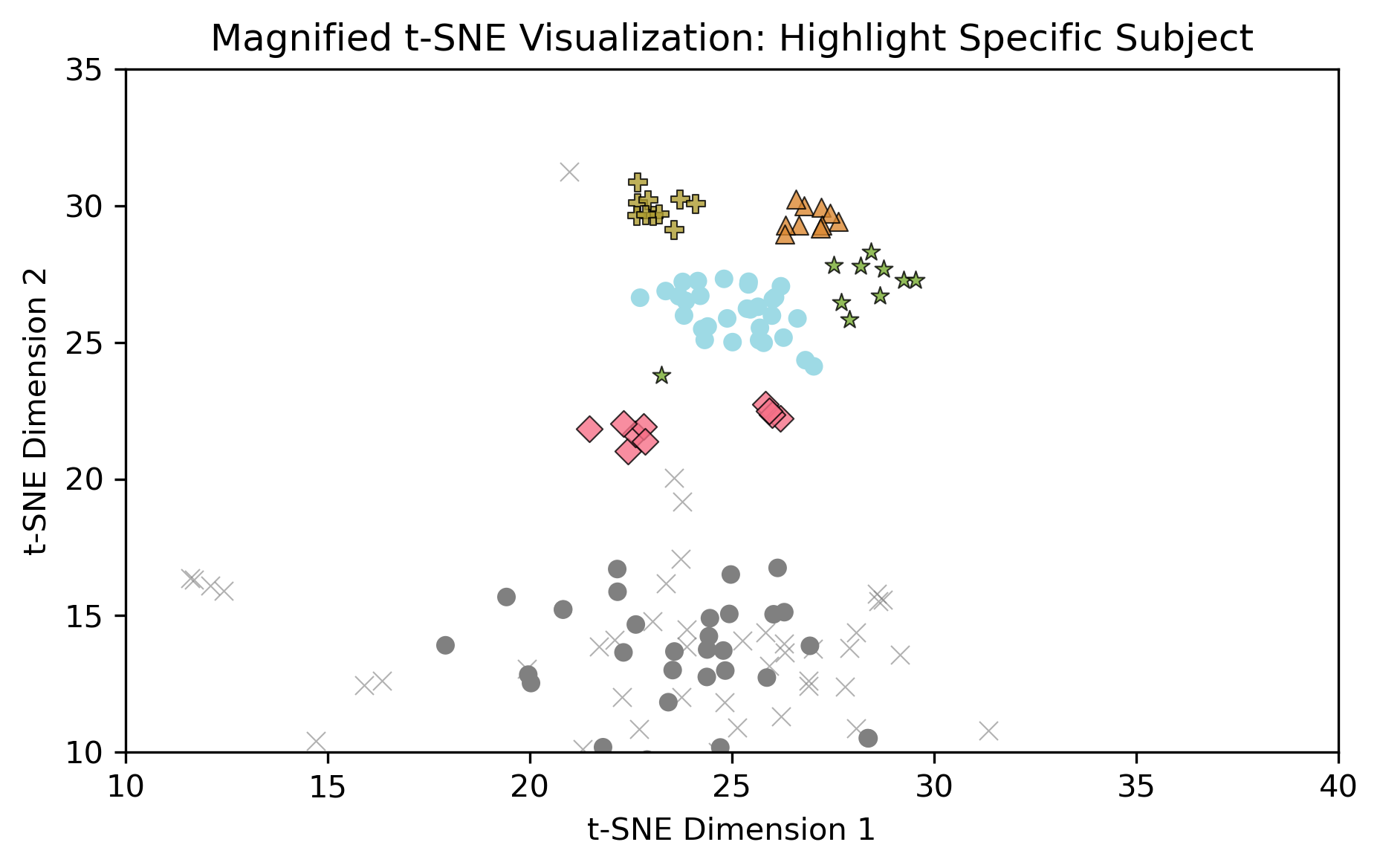}
    \caption{Magnified view of the upper-right region of Figure~\ref{fig:highl1}, highlighting the light blue subject. Each session for the highlighted subject is assigned a unique color, while all other subjects are shown in gray for context.}
    \label{fig:highl2}
\end{figure}

To investigate the embedding space while accounting for color and spatial limitations, we randomly selected 20 subjects to prevent overcrowding the embedding space with excessive data.\footnote{BioSemi subjects were excluded to avoid large gaps in the plot and enhance detail visibility.} Figure~\ref{fig:highl1} illustrates the embedding space for these selected subjects, demonstrating that the feature extraction method effectively identifies subject identities. However, a session effect is evident, where samples from the same session tend to form distinct clusters. To better visualize this phenomenon, the upper-right region of Figure~\ref{fig:highl1} is magnified in Figure~\ref{fig:highl2}, with separate colors indicating different sessions, further confirming the session-based clustering. The session effect on clustering may arise from environmental noise, changes in hair length, slight variations in the positioning of brainwave channels on the head, hardware differences, and natural changes in brainwave patterns over time due to factors such as aging, personal experiences, or mental states (e.g., stress or focus levels) throughout sessions. As a consequence, most samples within a session tend to be either close to or far from the enrollment set at the same time, which can contribute to high variability in performance, as reflected in the relatively large standard deviation ($\pm 6.3\%$) observed in the reported EER (10.28\%).

\subsubsection{Impact of Different Headsets on Performance}\footnote{To clarify, the results in IV.A (Table~II) and Sections IV.B.1–IV.B.3 are computed on the combined data from all the three headsets.}
\label{sec:headsets}
Differences in EEG recording hardware and channel positions require further investigation to understand their impact on brainwave authentication. We  set up to examine: (1) how well a model trained on one device performs on data from another, (2) whether models trained on multiple devices outperform device-specific models, and (3) how  performs varies when enrollment and verification data come from different devices.

\textbf{Cross Model:} To address the first question, we trained models using data from each specific headset and grouped the test data based on hardware to evaluate performance across models. The results, summarized in Table~\ref{tab:cross_headset_updated}, show that headset-specific models perform well only when tested on data from the same device. For example, the model trained on HydroCe achieves an EER of 9.2\% on HydroCe test data but 16.1\% on Geodesic test data. To further investigate why some hardware-specific models perform better on their own data, we summarized the training and testing data in Table~\ref{tab:dataset_summary2}, which reveals that the number of unique subjects in the training set has a stronger correlation with the EER. Therefore, weaker performance in some models could be related to the lower number of subjects available for learning features.

\textbf{General Model:} We trained a general model using data from all devices. The results in Table~\ref{tab:cross_headset_updated} show that the general model outperforms two hardware-specific models, even on their own data, but performs slightly worse than HydroCe. This comparison highlights two possible factors: the general model may benefit from learning cross-device patterns, or training on separate hardware-specific datasets may improve generalization. To examine this, we created a smaller training dataset by keeping only the most frequently used device per subject, ensuring each subject's data comes from a single device. The results in Table~\ref{tab:cross_headset2} show a similar pattern, although overall error rates are higher. Table~\ref{tab:dataset_summary2} also shows that HydroCe has significantly more training samples than other devices. These findings suggest that combining smaller datasets with consistent electrode positions can be beneficial for brainwave authentication, especially when large datasets are not available.

\textbf{Cross Device:} In real-world scenarios, individuals using brainwave-based authentication systems may rely on different recording devices. Therefore, authentication systems should maintain performance when enrollment and verification data come from different hardware.
%To evaluate this, we filtered genuine scores based on unique enrollment-verification device pairs and calculated the EER for each cross-device setting. 
To evaluate this, we used the general model trained on the combined training set from all three headsets (230 subjects) and then filtered genuine scores in the evaluation set (100 subjects) by enrollment–verification device pairs to compute the EER for each cross-device setting.
The results show an EER of 12.53\% for Geodesic–HydroCe pairs and 12.56\% for HydroCe–Geodesic pairs. In contrast, BioSemi–HydroCe and HydroCe–BioSemi pairs yield much higher EERs of 39.00\% and 44.05\%, respectively. For other combinations (e.g., Geodesic–BioSemi), no data were available in the evaluation set. The reason for this variation is that in the training set used for feature extraction, only 7 subjects had HydroCe–BioSemi session pairs and only 2 had BioSemi–HydroCe pairs out of 230 subjects, while 122 subjects had Geodesic–HydroCe session pairs. This limited overlap likely contributes to the poor generalization in the BioSemi–HydroCe condition and explains the formation of separate clusters in the t-SNE visualization shown in Figure~\ref{fig:SNE2}, where BioSemi forms a distinct cluster.

\begin{table}[h!]
\centering
\caption{EER (\%) and standard deviation (\(\pm\) Std) for different datasets/models. The first column (D) lists the datasets, while the first row (M) lists the models. For example, the entry in the second row and third column (28.32 \(\pm\) 11.7) represents the EER for the BioSemi model evaluated on the HydroCe dataset. "All" indicates results obtained using all data or models trained on data from all headsets.}
\begin{tabular}{lcccc}
\toprule
\textbf{D/M}      & \textbf{Geodisi}                  & \textbf{HydroCe}                  & \textbf{BioSemi}                & \textbf{All}                   \\
\midrule
\textbf{Geodisi}  & $9.52 \pm \text{\scriptsize 6.1}$  & $16.06 \pm \text{\scriptsize 8.4}$  & $29.33 \pm \text{\scriptsize 10.9}$ & $8.32 \pm \text{\scriptsize 5.2}$  \\
\textbf{HydroCe}  & $18.56 \pm \text{\scriptsize 10.0}$ & $9.24 \pm \text{\scriptsize 7.4}$  & $28.32 \pm \text{\scriptsize 11.7}$ & $9.90 \pm \text{\scriptsize 8.4}$  \\
\textbf{BioSemi}  & $32.63 \pm \text{\scriptsize 9.8}$  & $33.35 \pm \text{\scriptsize 10.5}$ & $23.52 \pm \text{\scriptsize 8.4}$  & $20.62 \pm \text{\scriptsize 7.6}$ \\
\textbf{All}      & $22.60 \pm \text{\scriptsize 12.6}$ & $22.08 \pm \text{\scriptsize 14.9}$ & $28.70 \pm \text{\scriptsize 11.6}$ & $10.28 \pm \text{\scriptsize 6.3}$ \\
\bottomrule
\end{tabular}
\label{tab:cross_headset_updated}
\end{table}

\begin{table}[h!]
\centering
\caption{Table showing EER (\%) with standard deviation (\(\pm\) Std). The first column (D) lists the datasets, and the first row (M) lists the models, where the training data for each subject is limited to their most frequently used hardware.}
\begin{tabular}{lcccc}
\toprule
\textbf{D/M}    & \textbf{Geodisi}             & \textbf{HydroCe}            & \textbf{BioSemi}            & \textbf{All}                \\
\midrule
\textbf{Geodisi}  & $17.37 \pm \text{\scriptsize 7.8}$  & $16.63 \pm \text{\scriptsize 8.6}$  & $30.10 \pm \text{\scriptsize 11.5}$ & $10.55 \pm \text{\scriptsize 6.3}$ \\
\textbf{HydroCe}  & $24.31 \pm \text{\scriptsize 9.6}$  & $10.91 \pm \text{\scriptsize 8.7}$  & $29.65 \pm \text{\scriptsize 12.2}$ & $10.09 \pm \text{\scriptsize 8.5}$ \\
\textbf{BioSemi}  & $33.52 \pm \text{\scriptsize 10.0}$ & $36.31 \pm \text{\scriptsize 11.2}$ & $22.71 \pm \text{\scriptsize 8.0}$  & $21.55 \pm \text{\scriptsize 7.9}$ \\
\textbf{All}      & $24.46 \pm \text{\scriptsize 11.5}$ & $23.27 \pm \text{\scriptsize 14.9}$ & $28.22 \pm \text{\scriptsize 13.4}$ & $21.09 \pm \text{\scriptsize 14.4}$ \\
\bottomrule
\end{tabular}
\label{tab:cross_headset2}
\end{table}

\begin{table}[h!]
\centering
\caption{Summary of training datasets by hardware type, listing sessions and subjects with percentages in parentheses.}
\begin{tabular}{@{}lcc@{}}
\toprule
\textbf{Hardware} & \textbf{Sessions} & \textbf{Subjects} \\
\midrule
\multicolumn{3}{l}{\textbf{Train Set: Full Dataset}} \\
\quad Hardware HydroCe    & 2038  & 169  \\
\quad Hardware Geodisi    & 1016  & 127  \\
\quad Hardware BioSemi    & 945   & 66   \\
\midrule
\multicolumn{3}{l}{\textbf{Train Set: Filtered Dataset}} \\
\quad Hardware HydroCe    & 1751   & 131   \\
\quad Hardware Geodisi    & 435    & 38   \\
\quad Hardware BioSemi    & 905   & 61   \\
\bottomrule
\end{tabular}
\label{tab:dataset_summary2}
\end{table}

\subsection{Enhancing Performance}
\label{sec:performance}
Samples from the same session are relatively close in the embedding space, but in some cases, they are more dispersed. For example, the pink session in Figure~\ref{fig:highl1}. Therefore utilizing multiple verification samples can improve results, albeit at the cost of additional verification time. Table~\ref{tab:avrage} illustrates the tradeoff between the number of verification samples, EER, and verification time. The results show that using more samples reduces the error, but improvements beyond four samples show a slower pace in reducing error. Considering the time cost of approximately one second per sample, using four samples appears to be a reasonable balance for the averaging strategy.

Moreover, while comparing the enrollment session to all verification sessions offers useful insights into brainwave authentication, real-world scenarios typically involve a single initial enrollment session followed by one verification attempt after some time. Upon successful login, it may be possible to update the enrollment set with new data from the verification attempt, which can then be used in future verifications. Therefore, in Table~\ref{tab:tradeoff}, we consider the first session as enrollment and the second as verification and report the results. We observe an EER of 6.87\% based on single verification sample and 3.04\% for four verification samples, showing significant improvement compared to the previous 10.28\% and 6.22\%. Additionally, we investigate the effect of multiple enrollment sessions. The results suggest that having two enrollment sessions improves EER to 4.87\% and 1.54\%, but further enrollment sessions show diminishing returns. Specifically, extending enrollment from two to three sessions reduces EER by only 0.08\,pp with 1V (4.87$\rightarrow$4.79) and slightly worsens with 4V (1.54$\rightarrow$1.57). While four sessions can lower EER (1V: 3.78; 4V: 1.14), these gains are not sustained at five and six sessions (1V: 4.42/4.99; 4V: 1.43/1.39). Thus, beyond two sessions the marginal benefit is small or unstable relative to the added user effort. The average time interval between the two enrollment sessions was 28 days. Among the subjects, 88 had a session interval of 10 days or less, and 32 had intervals of only one or two days.
\begin{table}[h!]
\centering
\caption{Tradeoff between the number of samples (SN) and EER. Increasing the number of samples reduces the EER but increases the verification time, as shown in the table.  }
\begin{tabular}{lcccccc}
\toprule
\textbf{SN} & 1   & 2   & 4   & 8    & 16    & 32   \\
\midrule
   EER     &10.28 & 7.75   & 6.22   & 5.35   & 4.87     & 4.59 \\
Time       & 1s  & 2s  & 4s  & 8s    & 16s    & 32s    \\
\bottomrule
\end{tabular}
\label{tab:avrage}
\end{table}

\begin{table}[h!]
\centering
\caption{Tradeoff between the number of enrollment sessions (ES) and EER when using 1 verification sample (1V) and 4 verification samples (4V). The table also includes the number of available subjects (NAS) used in the evaluation.}
\begin{tabular}{lcccccc}
\toprule
\textbf{ES} &  1   & 2   & 3   & 4   & 5    & 6   \\
\midrule
1V       &6.87 &  4.87   & 4.79   & 3.78   & 4.42   & 4.99 \\
4V       &3.04 &  1.54   & 1.57   & 1.14   & 1.43   & 1.39  \\
NAS       &100  &  99     & 99     & 98     & 91     & 75  \\
\bottomrule
\end{tabular}
\label{tab:tradeoff}
\end{table}
\begin{table}[h!]
\centering
\caption{EER (\%) when using the same channels as consumer-grade devices. \(E_x\) denotes the number of enrollment sessions, and \(xV\) denotes the number of verification samples. For the first two columns (Baseline and 4V), all possible verification sessions were considered. However, for the remaining columns, only the first verification session immediately following the last enrollment session was used.}
\begin{tabular}{lcccccc}
\toprule
\textbf{ES} &  Baseline  & 4V &  E1 1V  & E1 4V   & E2 1V    & E2 4V   \\
\midrule
Emotiv       &13.37 &8.68 &9.78   &4.85      & 7.57   &3.02  \\
DSI-VR300    &16.61 &11.55 &13.09   &7.11      &11.31   &5.35   \\
Muse 2       &18.18 &13.07 &15.01   &9.10     &12.74   &6.39   \\
\bottomrule
\end{tabular}
\label{tab:headsets}
\end{table}

\begin{table}[t]
\centering
\caption{Verification performance (EER [\%]) for different scalp regions (frontal, central, temporal, parietal–occipital)}
\begin{tabular}{lcccccc}
\toprule
\textbf{Region} & Baseline & 4V & E1 1V & E1 4V & E2 1V & E2 4V \\
\midrule
Frontal             & 15.74 & 10.36 & 12.51 &  6.49 & 10.63 &  5.13 \\
Central             & 16.36 & 11.31 & 12.41 &  6.59 & 10.67 &  4.94 \\
Temporal            & 15.76 & 10.90 & 11.72 &  6.48 & 10.11 &  4.53 \\
Par.--Occ. & 15.11 &  9.95 & 11.67 &  5.82 &  9.52 &  4.33 \\
\bottomrule
\end{tabular}
\label{tab:regions}
\end{table}

\begin{table*}[ht!]\centering
\caption{\small  Summary of state-of-the-art brain biometric verification solutions evaluated on multisession datasets}
\begin{tabular}{@{\extracolsep{2pt}}lcccccccc@{}}
\toprule

\multicolumn{1}{c}{Publication} & 
\multicolumn{1}{c}{Subjects} & 
\multicolumn{1}{c}{Sessions} &
\multicolumn{1}{c}{Interval} &
\multicolumn{1}{c}{Verification} &
\multicolumn{1}{c}{Dataset} &
\multicolumn{1}{c}{Open-source} &
\multicolumn{1}{c}{Inter-subject} &
\multicolumn{1}{c}{EER (\%)} 
\\

\cline{1-9}

				\textbf{Das \textit{et al.} ~\cite{das2016eeg}, 2016} 	
                & 50 & 150 (3) & up to 49 days& 30s &$\times$ & $\times$ & $\times$ &  $10 < \text{EER} < 22$

				\\\hline

				\textbf{Wu \textit{et al.} ~\cite{wu2018eeg}, 2018} 	
                & 40 & 80 (2) & 30 days& 6s &$\times$ & $\times$ & $\times$ &   $5.75 < \text{EER} < 7.07$
				\\\hline

				\textbf{Maiorana \textit{et al.} ~\cite{maiorana2017longitudinal}, 2018} 	
                & 45 & 270 (6) & up to 36 months& 5s &$\times$ & $\times$ & $\times$ &  $ 6.6  \leq \text{EER}  \leq 10.7 $

				\\\hline
				\textbf{Maiorana \cite{maiorana2021learning}, 2021}	
				& 45 & 225 (5) & up to 15 months& 5s &$\times$ & $\times$ & \checkmark & $4.8 \leq \text{EER} \leq 10.7$
								\\\hline

               \textbf{Debie \textit{et al.} \cite{debie2021session}, 2021}	
				& 54 & 108 (2) & different days\footnotemark[3]& $7.3s$ &\checkmark&  $\times$  & $\times$  &$1.16 < \text{EER} < 1.93$
								\\\hline

               \textbf{Chaurasia \textit{et al.} \cite{chaurasia2024neuroidbench}, 2024}	
				& 54 & 108 (2) & different days\footnotemark[3]& 1s &\checkmark  & \checkmark &\checkmark &  11.99
                \\\hline

               \textbf{Proposed Method}	
				& 345 & 6007 ($\sim 17$) & up to 66 months& 1s &\checkmark  & \checkmark & \checkmark & 10.28   
                \\\hline

								\bottomrule   

\end{tabular}

\label{tab:comparison}

\end{table*}

%\footnotemark[3]
\footnotetext[3]{The paper did not specify the exact number of days, so we used the same terminology. However, it typically refers to a very short time interval.}
\label{sec:consumer}
\subsection{Channels Reduction}
In this section, we analyze how reducing the number and placement of EEG channels affects biometric authentication performance. We consider two practically relevant scenarios: limiting channels to those available on specific consumer-grade headsets and restricting channels to subsets from distinct scalp regions under a fixed channel budget.

\subsubsection{Consumer-grade headsets}
Despite advancements in brainwave sensors in recent years, consumer-grade devices still have a limited number of sensors. Therefore, we selected three devices to evaluate performance by focusing exclusively on the electrodes specific to each device. The selected devices include the Emotiv Epoc X, which features 14\footnote{AF3, F7, F3, FC5, T7, P7, O1, O2, P8, T8, FC6, F4, F8, AF4} electrodes and costs 999 dollars, and the Muse 2, equipped with 4\footnote{TP9, AF7, AF8, TP10} electrodes and available at 269.99 euros. Both are budget-friendly options for consumer use. Additionally, we considered the DSI-VR300\footnote{FCz, Pz, P3, P4, PO7, PO8, Oz}, which includes 7 electrodes and costs  10,000 dollars. This device, though expensive, has the potential to become more affordable in the future and could be integrated into XR devices. We simulated consumer-grade devices by selecting, from the PEERS dataset, the electrodes corresponding to each device’s channel layout. We subsequently trained a separate model for each configuration, following the same subject-disjoint split (230 subjects for training, 100 for evaluation). Table VIII represents the results with a lower number of channels. The baseline EER increases from 10.28\% to 13.37\%, 16.61\%, and 18.18\% for Emotiv Epoc X, DSI-VR300, and Muse 2, respectively.  The reduction from 93 channels in our main dataset to 14, 7, and 4 channels suggests redundancy due to correlated EEG signals across electrodes. Moreover, using averaging techniques and multiple enrollment sessions can further improve the results.

\subsubsection{Regional channel subsets}
To analyze which scalp regions contribute most to authentication performance under a fixed channel budget, we grouped electrodes into frontal, central, temporal, and parietal–occipital regions according to their 10–20 labels. We first counted, in each region, how many such symmetric pairs were available and found that the smallest region contained six pairs; we therefore fixed the subset size to six pairs (12 channels) in all regions so that every regional model operated with the same number and type of electrodes. Separate models were trained on each regional subset using the same architecture and protocol as in the full-channel setting, and the resulting EERs are reported in Table~\ref{tab:regions}. The parietal–occipital subset achieved the lowest error rates, followed by frontal and central subsets, while temporal channels yielded slightly higher EERs. This indicates that placing electrodes over parietal–occipital areas preserves authentication performance best, in line with previous EEG biometric studies\cite{maiorana2021learning,la2014human}. %This region is also less affected by eye-movement artifacts, which mainly disturb electrodes on the forehead close to the eyes\cite{ronca2024preprocessing}. Therefore, parietal–occipital placements should be considered when designing consumer-grade devices, as they combine high discriminative power with reduced sensitivity to ocular artifacts.

Overall, these channel-reduction experiments indicate that a subset of electrodes can retain much of the authentication performance while improving usability. Our consumer-grade headset configurations show that a substantial fraction of the baseline performance can be preserved even with only 4--14 channels, suggesting that low-channel-count devices can be effective when properly configured. Regarding electrode placement, parietal–occipital sites demonstrate the best performance, which may be partly explained by the fact that they are less affected by eye-movement artifacts that mainly disturb electrodes on the forehead close to the eyes~\cite{ronca2024preprocessing}. Therefore, these sites are particularly attractive for future consumer-grade devices.Moreover, as discussed in Sections~IV-B3 and~IV-B4, hardware characteristics affect performance, but this can be mitigated by providing sufficient data from target devices and incorporating such data into model training, which will be crucial for achieving reliable results. Finally, if developers adopt a common set of electrode locations, this would support data pooling across devices and studies, benefiting both authentication and other EEG-based applications.

%Overall, these channel-reduction experiments indicate that a subset of electrodes can retain much of the authentication performance while improving usability. Parietal–occipital placements are also less affected by eye-movement artifacts, which mainly disturb electrodes on the forehead close to the eyes~\cite{ronca2024preprocessing}; therefore, they combine high discriminative power with reduced sensitivity to ocular artifacts and are particularly attractive for future consumer-grade devices. At the same time, our consumer-grade headset configurations show that a substantial fraction of the baseline performance can be preserved even with only 4--14 channels, suggesting that low-channel-count devices can be effective when properly configured. As discussed in Sections~IV-B3 and~IV-B4, hardware characteristics and data availability have a measurable impact on performance; providing sufficient data from target devices and contributing such data to the training sets of public models will be crucial for achieving reliable results. Moreover, if developers converge on a small set of common electrode locations, this would facilitate pooling data across devices and studies, which could benefit both authentication and other EEG-based applications.

%We selected the same electrodes as used in the devices and trained a model for each device. 

\subsection{Comparative Analysis}
\label{sec:comparative}
We first evaluate the proposed method against state-of-the-art brainwave authentication approaches. Next, we analyze its performance relative to international biometric standards to provide context and clarify the significance of the results.

\subsubsection{Comparison with Brainwave Authentications Systems}
Table~\ref{tab:comparison} summarizes the state-of-the-art in multi-session brainwave authentication studies.

\textbf{Das et al.~\cite{das2016eeg}} is one of the earliest multi-session studies, where the authors used simple averaging of samples and compared them using cosine distance. Their reported EER ranges between approximately 10\% and 22\%. The best result for a one-week interval is around 10\% EER, while for an average interval of 34 days, the EER is approximately 15\%.

\textbf{Wu et al.\cite{wu2018eeg} and Debie et al.\cite{debie2021session}} do not report EER directly. Wu et al. report an FRR of 7.07\% and FAR of 5.75\% with a 6-second window, while Debie et al. report an FRR of 1.93\% and FAR of 1.16\% using 7.3 seconds of data with only a few days between sessions. Debie et al.\cite{debie2021session} performance is calculated by averaging two scenarios: (1) first session as enrollment and second as verification, and (2) the reverse---an unrealistic setup that assumes access to future data. Both studies use two-session datasets and train a separate CNN per subject, making it unclear how session variability is learned. This design limits generalization and requires training a new model for each user, which is impractical and does not support inter-subject learning.
 
\textbf{Maiorana et al.~\cite{maiorana2017longitudinal, maiorana2021learning}} present a longitudinal evaluation with 45 subjects over six sessions collected across three years. In their first study, they used handcrafted features with a Hidden Markov Model and reported EERs between 6.6\% and 10.7\% for 5-second verification samples. In the second study, the evaluation was limited to 15 months without a clear explanation. They adopted a metric learning approach, training a feature extractor using contrastive loss on 30 subjects, with a separate CNN per channel, and used a one-class SVM for classification. They reported EERs between 4.8\% and 10.7\% based on 15 evaluation subjects. However, this approach has key limitations: learning from single channels ignores inter-channel relationships, and one-class SVMs have shown limited reliability, as supported by our findings.

\textbf{Chaurasia et al.\cite{chaurasia2024neuroidbench}} present a framework for EEG evaluation that investigates the multi-session scenario. They show that, on a small dataset, handcrafted features outperform deep learning approaches %in the multi-session setting, 
, reporting an EER of 11.99\% using PSD and AR features combined with an SVM, compared to 20.68\% EER using a CNN with triplet loss.

In terms of results, when using the first session for enrollment and the second for verification---with an average interval of 28 days---we achieve an EER of 3.04\%, outperforming \cite{das2016eeg,wu2018eeg,chaurasia2024neuroidbench}. Maiorana et al.\cite{maiorana2017longitudinal,maiorana2021learning} report 4.8\% EER with a 5-second window and similar time interval (a month), while we obtain 3.04\% with a 4-second window. Compared to Debie et al.~\cite{debie2021session}, who report EER between 1.93\% and 1.16\% using a 7.3-second window and only a few days between sessions, our 2.30\% EER with a 7-second window is measured across a average of 28-day interval. Overall, our results either outperform or are competitive with the state of the art.

However, several important issues are overlooked in related works: (1) Most of the studies use short intervals~\cite{das2016eeg,wu2018eeg,debie2021session,wu2018eeg} between sessions while our work considers longer intervals, making the evaluation more realistic for practical applications. (2) Except for Chaurasia et al.\cite{chaurasia2024neuroidbench}, none of the studies provide source code, making it impossible to reproduce results or build upon them. This lack of transparency makes it particularly difficult to verify findings, especially where reported performance differs from our results and expectations, such as the use of one-class SVM~\cite{maiorana2021learning} and subject-specific CNNs~\cite{debie2021session,wu2018eeg}  (3) None of the reviewed studies examine the effect of recording hardware on authentication performance, which is a key factor for real-world deployment. (4) Some related works~\cite{das2016eeg,wu2018eeg,debie2021session,maiorana2017longitudinal} do not follow an inter-subject approach, where a general feature extractor is trained to avoid the need for handcrafted features or retraining a new network for each new subject---similar to best practices in face recognition systems~\cite{deng2019arcface,schroff2015facenet,khosla2020supervised}. (5) All studies use only one session for verification, even when multiple are available. They typically repeat experiments with different session pairs (one enrollment - one verification), averaging the results. In contrast, using the first session for enrollment and the rest for verification reflects a more realistic deployment setting.

\begin{table*}[ht]
\centering
\caption{Performance metrics for different scenarios with detailed enrollment and verification session information. (Verification samples refer to the number of samples from the same session used in a single verification attempt.) }
\begin{tabular}{lllcccc}
\toprule
\multicolumn{3}{c}{Scenario} & \multicolumn{4}{c}{Metrics (\%)} \\
\cmidrule(r){1-3} \cmidrule(l){4-7}
Enrollment Sess. & Verification Sess. & Verification Samples & EER & FMR 1\% & FMR 0.1\% & FMR 0.01\% \\
\midrule
First & All Others & 1 & $10.28 \pm 6.30$ & $43.33 \pm 24.46$ & $66.98 \pm 25.30$ & $80.65 \pm 21.45$ \\
First & All Others  & 4 & $6.22 \pm 5.64$  & $27.86 \pm 25.75$ & $50.39 \pm 30.40$ & $64.75 \pm 30.03$ \\
First & Second & 1 & $6.87 \pm 7.61$ & $32.15 \pm 32.29$ & $54.04 \pm 35.66$ & $65.07 \pm 34.52$ \\
First & Second & 4 & $3.04 \pm 6.23$ & $20.71 \pm 33.98$ & $33.44 \pm 40.58$ & $41.26 \pm 42.16$ \\

First Two & Third & 1 & $4.87 \pm 5.36$ & $23.98 \pm 28.29$ & $43.59 \pm 35.04$ & $56.39 \pm 35.15$\\
First Two & Third& 4  & $1.54 \pm 2.74$ & $12.24 \pm 26.99$ & $24.28 \pm 35.33$ & $32.11 \pm 38.79$\\
First Two & Third& 16  & $0.68 \pm 1.75$ & $9.60 \pm 26.67$ & $16.08 \pm 33.37$ & $19.99 \pm 36.31$\\

\bottomrule
\end{tabular}
\label{tab:detailed_metrics}
\end{table*}

\subsubsection{Comparison with Biometric Industry Standards}
Brainwave authentication is in its early stages, making it crucial to assess progress and identify gaps for real-world implementation. For a biometric system to be deployed in practical applications, it must adhere to international industrial biometric standards. 
These standards emphasize a very low FAR to ensure security, while maintaining a reasonable FRR to keep the system usable. %for legitimate users.
 According to guidelines from NIST (2023)/ISO\footnote{\scriptsize\url{https://pages.nist.gov/800-63-3/sp800-63b.html}}and the European Border Guard Agency Frontex\footnote{\scriptsize\url{https://www.frontex.europa.eu/assets/Publications/Research/Best_Practice_Technical_Guidelines_ABC.pdf}}, biometric systems %are required to 
 should operate at FAR $\leq 0.1\%$ (1 in 1,000). Stricter recommendations from FIDO\footnote{\scriptsize\url{https://fidoalliance.org/specs/biometric/requirements/Biometrics-Requirements-v4.0.1-fd-20240522.pdf}} and more recent NIST (August 2024)/ISO 2024 standards\footnote{\scriptsize\url{https://pages.nist.gov/800-63-4/sp800-63b.html}} suggest a FAR $\leq 0.01\%$ (1 in 10,000). For usability, FRR standards propose a maximum FRR of $\leq 5\%$, allowing at least 19 successful logins out of 20 attempts for genuine users. Table~\ref{tab:detailed_metrics} presents the FRR at low FAR of 1\%, 0.1\%, and 0.01\%. The results show higher FRR. For instance, using the first session as enrollment and all others as verification, with a duration of 4 seconds, the FRR is 50.39\% at FAR 0.1\% and 64.75\% at FAR 0.01\%. This means approximately 9 and 7 successful logins out of 20 attempts for legitimate users, respectively.

\begin{figure}
    \centering
    \includegraphics[width=0.8\linewidth]{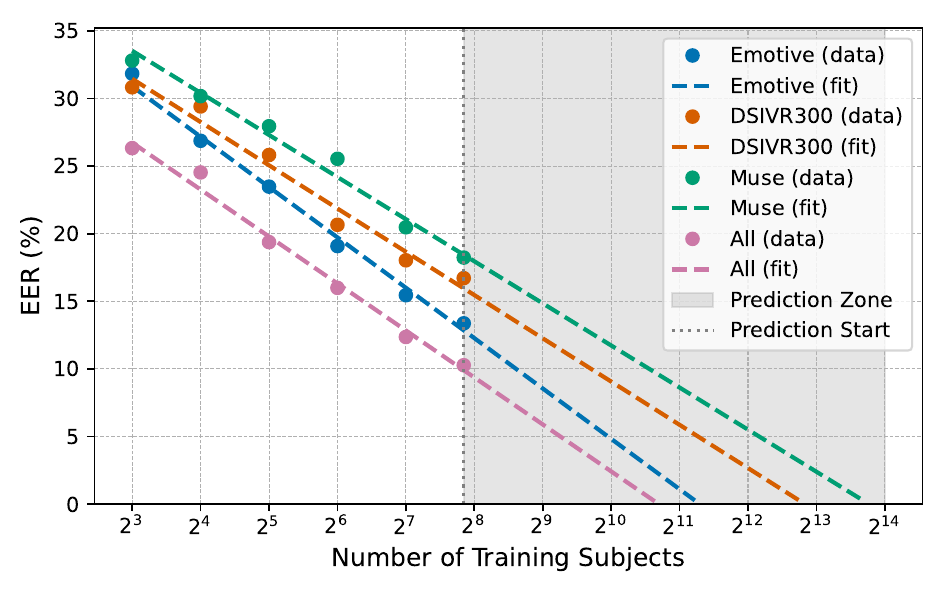}

    \caption{EER vs. number of training subjects. Points show experimental results; dashed line shows linear prediction.}
    \label{fig:eer-subjectn}
\end{figure}

While our experiments show that brainwave authentication does not yet meet international standards, our previous experiment on the impact of headsets on performance (Sec.\ref{sec:headsets}) revealed that the headset with a higher number of subjects yielded a lower error rate, and we also know that biometrics that meet these standards typically employ datasets with thousands of subjects\cite{Guo2016MSCeleb1MAD}. Therefore, we conducted an experiment to explore the relationship between the number of subjects used for training the feature extractor and the error rate. The results in Figure~\ref{fig:eer-subjectn} indicate a linear relationship between the logarithm of the number of subjects and the EER. It is obvious that the linear decline eventually slows down. The current trend suggests that with a dataset of a few thousand subjects (from 2,000 to 16,000, depending on the device), it may be possible to approach biometric performance standards. This is reasonable when compared to the data volumes typically used in modalities like face recognition.

\section{Limitations}

Our study is subject to certain limitations at three levels: the authentication task, the dataset size, and the recoding hardware.

%At the authentication task level, our stimuli were limited to a general text-based memorization task. Ideally, multiple task types could be included to allow for performance comparisons and identification of the best-performing task. Additionally, we did not explicitly address replay attacks or EEG-based liveness detection. These aspects can be mitigated by adapting the approach of Lin et al.~\cite{lin2018brain}, which involves varying the stimulus and detecting corresponding changes in EEG responses.

At the authentication task level, our stimuli were limited to a general text-based memorization task, which imposes non-negligible cognitive load, requires explicit focused attention during authentication, and may limit usability in practical settings; therefore, the usability of this specific task requires further investigation, despite existing general usability studies on brainwave authentication~\cite{fallahi2024usability}. Ideally, multiple task types could be included to allow for performance comparisons and identification of the best-performing task. Additionally, we did not explicitly address replay attacks or EEG-based liveness detection. These aspects can be mitigated by adapting the approach of Lin et al.~\cite{lin2018brain}, which involves varying the stimulus and detecting corresponding changes in EEG responses.

At the dataset size level, we only had access to a single dataset including a notable number of sessions and subjects. However, best practices in state-of-the-art face recognition research recommend training on multiple large datasets and testing on an independent dataset to ensure a more robust evaluation~\cite{kim2022adaface}. To simulate this, we split our dataset into a subject-independent set for training feature extraction and evaluation. Moreover, we partitioned the dataset based on device, showing that cross-device evaluation significantly increases error (Tables~\ref{tab:cross_headset_updated} and~\ref{tab:cross_headset2}.) Achieving robust performance requires large amounts of data to extract distinctive features and account for subject-specific, environmental, and hardware-related variations. For example, in the case of face recognition, Kim et al.~\cite{kim2022adaface} used 3 datasets comprising over 350,000 subjects and more than 14.9 million images to train a generalized model. Such large-scale datasets are not yet available in brainwave authentication.

At the hardware level, we simulated consumer-grade devices by selecting the same number and channel locations as those found in such devices. However, we did not have a measure to directly compare the quality of sensors. Although sensor quality will improve, the main challenge will remain their number and placement for effective use. It is worth noting that all other multi-session studies in the literature~\cite{maiorana2017longitudinal,debie2021session,wu2018eeg,chaurasia2024neuroidbench} also employ medical-grade data.

\section{Conclusion and Future Work}
This study  investigated  factors influencing the performance of brainwave authentication over multiple sessions. Our findings highlight that the pairwise deep learning approach for feature extraction outperforms hand-crafted features. Interestingly, we observed that larger evaluation sets do not necessarily make authentication more difficult; instead, they reduce the uncertainty of the results. 
We acknowledge that the error rate will eventually increase as the number of individuals grows, since only a limited number of individual templates can be reliably distinguished within a finite embedding space. Expanding the embedding space could mitigate this limitation, though evaluating this would require datasets with substantially more users.

Authentication performance remained relatively stable during the first week, with error rates increasing only slightly in the second month. However, error rates could increase by up to 2.1 times after one year, suggesting that the enrollment dataset should be periodically updated to maintain performance. We also examined the impact of recording devices, emphasizing the importance of including data from different headsets in training the feature extractor. Additionally, we showed that increasing the number of enrollment sessions and verification time can help reduce the EER, though only up to a certain point.  Our findings also demonstrate that brainwave authentication remains feasible even with a reduced number of EEG sensors. Finally,  our results are competitive with the state-of-the-art in brainwave authentication, investigated with stronger assumptions in simpler scenarios.

For future work, there is an urgent need for a public multi-session dataset with a few thousand subjects to ensure brainwave authentication research meets international standards. This dataset should be collected using consumer-grade devices in multiple sessions, and incorporating multiple devices in the same session would be valuable for studying the cross-headset authentication challenge, and across different task paradigms (e.g., passive viewing, motor tasks) to study task similarity. On the technical side, further research should focus on developing deep learning architectures that better capture the unique features of brainwave data, as well as effective signal preprocessing approaches to remove noise. Specifically, the new trend of utilizing attention mechanisms and specialized transformer architectures~\cite{nguyen2024spatio}, which have shown promising results in other biometric modalities (e.g., enhanced keystroke dynamics~\cite{senarath2023behaveformer}), needs future investigation to see how they can be effectively integrated into EEG-based systems for improved feature extraction and long-term performance. Finally, adopting an open-source approach would enable researchers to build on each other's work, promoting reproducibility and collaboration. Further investigation of more advanced attacks on brainwave authentication is necessary.

%For future work, there is an urgent need for a public multi-session dataset with a few thousand subjects to ensure brainwave authentication research meets international standards. This dataset should be collected using consumer-grade devices in multiple sessions, and incorporating multiple devices in the same session would be valuable for studying the cross-headset authentication challenge. On the technical side, further research should focus on developing deep learning architectures that better capture the unique features of brainwave data, as well as effective signal preprocessing approaches to remove noise. Specifically, the new trend of utilizing attention mechanisms and specialized transformer architectures~\cite{nguyen2024spatio}, which have shown promising results in other biometric modalities (e.g., enhanced keystroke dynamics~\cite{senarath2023behaveformer}), needs future investigation to see how they can be effectively integrated into EEG-based systems for improved feature extraction and long-term performance. Finally, adopting an open-source approach would enable researchers to build on each other's work, promoting reproducibility and collaboration. Further investigation of more advanced attacks on brainwave authentication is necessary.

\section*{Acknowledgment}
This work was funded by the Topic Engineering Secure Systems of the Helmholtz Association (HGF) and supported by KASTEL Security Research Labs, Karlsruhe, and Germany’s Excellence Strategy (EXC 2050/1 ‘CeTI’; ID 390696704).

\bibliographystyle{IEEEtran}
\bibliography{refrences}

@article{kahana2024penn,
  title={The Penn Electrophysiology of Encoding and Retrieval Study.},
  author={Kahana, Michael J and Lohnas, Lynn J and Healey, M Karl and Aka, Ada and Broitman, Adam W and Crutchley, Patrick and Crutchley, Elizabeth and Alm, Kylie H and Katerman, Brandon S and Miller, Nicole E and others},
  journal={Journal of Experimental Psychology: Learning, Memory, and Cognition},
  year={2024},
  publisher={American Psychological Association}
}

@inproceedings{senarath2023behaveformer,
  title={Behaveformer: A framework with spatio-temporal dual attention transformers for imu-enhanced keystroke dynamics},
  author={Senarath, Dilshan and Tharinda, Sanuja and Vishvajith, Maduka and Rasnayaka, Sanka and Wickramanayake, Sandareka and Meedeniya, Dulani},
  booktitle={2023 IEEE International Joint Conference on Biometrics (IJCB)},
  pages={1--9},
  year={2023},
  organization={IEEE}
}

@article{ronca2024preprocessing,
  title={Optimizing EEG Signal Integrity: A Comprehensive Guide to Ocular Artifact Correction},
  author={Ronca, Vincenzo and Capotorto, Rossella and Di Flumeri, Gianluca and Giorgi, Andrea and Vozzi, Alessia and Germano, Daniele and Virgilio, Valerio Di and Borghini, Gianluca and Cartocci, Giulia and Rossi, Dario and others},
  journal={Bioengineering},
  volume={11},
  number={10},
  pages={1018},
  year={2024},
  publisher={MDPI}
}

@article{la2014human,
  title={Human brain distinctiveness based on EEG spectral coherence connectivity},
  author={La Rocca, Daria and Campisi, Patrizio and Vegso, Balazs and Cserti, Peter and Kozmann, Gy{\"o}rgy and Babiloni, Fabio and Fallani, F De Vico},
  journal={IEEE transactions on Biomedical Engineering},
  volume={61},
  number={9},
  pages={2406--2412},
  year={2014},
  publisher={IEEE}
}

@article{nguyen2024spatio,
  title={Spatio-temporal dual-attention transformer for time-series behavioral biometrics},
  author={Nguyen, Kim-Ngan and Rasnayaka, Sanka and Wickramanayake, Sandareka and Meedeniya, Dulani and Saha, Sanjay and Sim, Terence},
  journal={IEEE Transactions on Biometrics, Behavior, and Identity Science},
  volume={6},
  number={4},
  pages={591--601},
  year={2024},
  publisher={IEEE}
}

@article{lawhern2018eegnet,
  title={EEGNet: a compact convolutional neural network for EEG-based brain--computer interfaces},
  author={Lawhern, Vernon J and Solon, Amelia J and Waytowich, Nicholas R and Gordon, Stephen M and Hung, Chou P and Lance, Brent J},
  journal={Journal of neural engineering},
  volume={15},
  number={5},
  pages={056013},
  year={2018},
  publisher={iOP Publishing}
}

@inproceedings{wang2019multi, title={Multi-similarity loss with general pair weighting for deep metric learning}, author={Wang, Xun and Han, Xintong and Huang, Weilin and Dong, Dengke and Scott, Matthew R}, booktitle={Proceedings of the IEEE/CVF conference on computer vision and pattern recognition}, pages={5022--5030}, year={2019} }

@inproceedings{he2016deep,
  title={Deep residual learning for image recognition},
  author={He, Kaiming and Zhang, Xiangyu and Ren, Shaoqing and Sun, Jian},
  booktitle={Proceedings of the IEEE conference on computer vision and pattern recognition},
  pages={770--778},
  year={2016}
}

@inproceedings{ang2008filter,
  title={Filter bank common spatial pattern (FBCSP) in brain-computer interface},
  author={Ang, Kai Keng and Chin, Zheng Yang and Zhang, Haihong and Guan, Cuntai},
  booktitle={2008 IEEE international joint conference on neural networks (IEEE world congress on computational intelligence)},
  pages={2390--2397},
  year={2008},
  organization={IEEE}
}

@article{simonyan2014very,
  title={Very deep convolutional networks for large-scale image recognition},
  author={Simonyan, Karen and Zisserman, Andrew},
  journal={arXiv preprint arXiv:1409.1556},
  year={2014}
}

@misc{suppiah2018biometric,
  title={Biometric identification using single channel EEG during relaxed resting state. IET Biometr 7: 342--348},
  author={Suppiah, R and Prasad Vinod, A},
  year={2018}
}

@inproceedings{zheng2022task,
  title={Task-oriented self-supervised learning for anomaly detection in electroencephalography},
  author={Zheng, Yaojia and Liu, Zhouwu and Mo, Rong and Chen, Ziyi and Zheng, Wei-shi and Wang, Ruixuan},
  booktitle={International Conference on Medical Image Computing and Computer-Assisted Intervention},
  pages={193--203},
  year={2022},
  organization={Springer}
}

@book{zhang2021deep,
  title={Deep learning for EEG-based brain--computer interfaces: Representations, algorithms and applications},
  author={Zhang, Xiang and Yao, Lina},
  year={2021},
  publisher={World Scientific}
}

@article{schirrmeister2017deep,
  title={Deep learning with convolutional neural networks for EEG decoding and visualization},
  author={Schirrmeister, Robin Tibor and Springenberg, Jost Tobias and Fiederer, Lukas Dominique Josef and Glasstetter, Martin and Eggensperger, Katharina and Tangermann, Michael and Hutter, Frank and Burgard, Wolfram and Ball, Tonio},
  journal={Human brain mapping},
  volume={38},
  number={11},
  pages={5391--5420},
  year={2017},
  publisher={Wiley Online Library}
}

@inproceedings{nai2004classification,
  title={Classification of mental tasks using fixed and adaptive autoregressive models of EEG signals},
  author={Nai-Jen, Huan and Palaniappan, Ramaswamy},
  booktitle={The 26th annual international conference of the IEEE engineering in medicine and biology society},
  volume={1},
  pages={507--510},
  year={2004},
  organization={IEEE}
}

@article{gramfort2014mne,
  title={MNE software for processing MEG and EEG data},
  author={Gramfort, Alexandre and Luessi, Martin and Larson, Eric and Engemann, Denis A and Strohmeier, Daniel and Brodbeck, Christian and Parkkonen, Lauri and H{\"a}m{\"a}l{\"a}inen, Matti S},
  journal={neuroimage},
  volume={86},
  pages={446--460},
  year={2014},
  publisher={Elsevier}
}

@article{debie2021session,
  title={Session invariant EEG signatures using elicitation protocol fusion and convolutional neural network},
  author={Debie, Essam and Moustafa, Nour and Vasilakos, Athanasios},
  journal={IEEE Transactions on Dependable and Secure Computing},
  volume={19},
  number={4},
  pages={2488--2500},
  year={2021},
  publisher={IEEE}
}

@article{mansfield2002best,
  title={Best practices in testing and reporting performance of biometric devices},
  author={Mansfield, Anthony J and Wayman, James L},
  year={2002},
  publisher={Centre for Mathematics and Scientific Computing, National Physical~…}
}

@inproceedings{kim2022adaface,
  title={Adaface: Quality adaptive margin for face recognition},
  author={Kim, Minchul and Jain, Anil K and Liu, Xiaoming},
  booktitle={Proceedings of the IEEE/CVF conference on computer vision and pattern recognition},
  pages={18750--18759},
  year={2022}
}

@article{haas2003hans,
  title={Hans berger (1873--1941), richard caton (1842--1926), and electroencephalography},
  author={Haas, Lindsay F},
  journal={Journal of Neurology, Neurosurgery \& Psychiatry},
  volume={74},
  number={1},
  pages={9--9},
  year={2003},
  publisher={BMJ Publishing Group Ltd}
}

@book{niedermeyer2005electroencephalography, title={Electroencephalography: basic principles, clinical applications, and related fields}, author={Niedermeyer, Ernst and da Silva, FH Lopes}, year={2005}, publisher={Lippincott Williams \& Wilkins} }

@book{adeli2010automated,
  title={Automated EEG-based diagnosis of neurological disorders: Inventing the future of neurology},
  author={Adeli, Hojjat and Ghosh-Dastidar, Samanwoy},
  year={2010},
  publisher={CRC press}
}

@article{welch1967,
  author = {Welch, Peter D.},
  title = {The Use of Fast Fourier Transform for the Estimation of Power Spectra: A Method Based on Time Averaging Over Short, Modified Periodograms},
  journal = {IEEE Transactions on Audio and Electroacoustics},
  volume = {15},
  number = {2},
  pages = {70--73},
  year = {1967},
  publisher = {IEEE}
}

@article{ayeswarya2024comprehensive,
  title={A comprehensive review on secure biometric-based continuous authentication and user profiling},
  author={Ayeswarya, S and Singh, K John},
  journal={IEEE Access},
  year={2024},
  publisher={IEEE}
}

@inproceedings{oh2016deep,
  title={Deep metric learning via lifted structured feature embedding},
  author={Oh Song, Hyun and Xiang, Yu and Jegelka, Stefanie and Savarese, Silvio},
  booktitle={Proceedings of the IEEE conference on computer vision and pattern recognition},
  pages={4004--4012},
  year={2016}
}

@article{jiang2023cancelable,
  title={Cancelable biometric schemes for Euclidean metric and Cosine metric},
  author={Jiang, Yubing and Shen, Peisong and Zeng, Li and Zhu, Xiaojie and Jiang, Di and Chen, Chi},
  journal={Cybersecurity},
  volume={6},
  number={1},
  pages={4},
  year={2023},
  publisher={Springer}
}

@article{khosla2020supervised,
  title={Supervised contrastive learning},
  author={Khosla, Prannay and Teterwak, Piotr and Wang, Chen and Sarna, Aaron and Tian, Yonglong and Isola, Phillip and Maschinot, Aaron and Liu, Ce and Krishnan, Dilip},
  journal={Advances in neural information processing systems},
  volume={33},
  pages={18661--18673},
  year={2020}
}

@inproceedings{deng2019arcface,
  title={Arcface: Additive angular margin loss for deep face recognition},
  author={Deng, Jiankang and Guo, Jia and Xue, Niannan and Zafeiriou, Stefanos},
  booktitle={Proc. of the IEEE/CVF conf. on computer vision and pattern recognition},
  pages={4690--4699},
  year={2019}
}

@inproceedings{Guo2016MSCeleb1MAD,
    author = "Guo, Yandong and Zhang, Lei and Hu, Yuxiao and He, X. and Gao, Jianfeng",
    title = "MS-Celeb-1M: A Dataset and Benchmark for Large-Scale Face Recognition",
    booktitle = "ECCV",
    year = "2016"
}

@article{horiguchi2019significance,
  title={Significance of softmax-based features in comparison to distance metric learning-based features},
  author={Horiguchi, Shota and Ikami, Daiki and Aizawa, Kiyoharu},
  journal={IEEE Trans. on pattern analysis and machine intelligence},
  volume={42},
  number={5},
  pages={1279--1285},
  year={2019},
  publisher={IEEE}
}

@article{kulis2013metric,
  title={Metric learning: A survey},
  author={Kulis, Brian and others},
  journal={Foundations and Trends{\textregistered} in Machine Learning},
  volume={5},
  number={4},
  pages={287--364},
  year={2013},
  publisher={Now Publishers, Inc.}
}

@inproceedings{schroff2015facenet,
  title={Facenet: A unified embedding for face recognition and clustering},
  author={Schroff, Florian and Kalenichenko, Dmitry and Philbin, James},
  booktitle={Proc.s of the IEEE conference on computer vision and pattern recognition},
  pages={815--823},
  year={2015}
}

@article{khalil2024unlocking,
  title={Unlocking Security for Comprehensive Electroencephalogram-Based User Authentication Systems},
  author={Khalil, Adnan Elahi Khan and Perez-Diaz, Jesus Arturo and Cantoral-Ceballos, Jose Antonio and Antelis, Javier M},
  journal={Sensors},
  volume={24},
  number={24},
  pages={7919},
  year={2024},
  publisher={MDPI}
}

@article{raurale2021emg, title={EMG biometric systems based on different wrist-hand movements}, author={Raurale, Sumit A and McAllister, John and Del Rinc{\'o}n, Jes{\'u}s Mart{\'\i}nez}, journal={IEEE Access}, volume={9}, pages={12256--12266}, year={2021}, publisher={IEEE} }

@inproceedings{nath2023acute,
  title={Acute stress data-based fast biometric system using contrastive learning and ultra-short ecg signal segments},
  author={Nath, Rajdeep K and Tervonen, Jaakko and N{\"a}rv{\"a}inen, Johanna and Pettersson, Kati and M{\"a}ntyj{\"a}rvi, Jani},
  booktitle={Adjunct Proceedings of the 2023 ACM International Joint Conference on Pervasive and Ubiquitous Computing \& the 2023 ACM International Symposium on Wearable Computing},
  pages={642--647},
  year={2023}
}

@article{peng2019eeg,
  title={EEG preprocessing and denoising},
  author={Peng, Weiwei},
  journal={EEG Signal Processing and Feature Extraction},
  pages={71--87},
  year={2019},
  publisher={Springer}
}

@inproceedings{fallahi2024usability,
  title={On the usability of next-generation authentication: A study on eye movement and brainwave-based mechanisms},
  author={Fallahi, Matin and Arias-Cabarcos, Patricia and Strufe, Thorsten},
  booktitle={Proceedings of the Extended Abstracts of the CHI Conference on Human Factors in Computing Systems},
  year={2025}
}

@article{silverman1963rationale,
  title={The rationale and history of the 10-20 system of the International Federation},
  author={Silverman, Daniel},
  journal={American Journal of EEG Technology},
  volume={3},
  number={1},
  pages={17--22},
  year={1963},
  publisher={Taylor \& Francis}
}

@article{wang2022cancellable,
  title={Cancellable template design for privacy-preserving EEG biometric authentication systems},
  author={Wang, Min and Wang, Song and Hu, Jiankun},
  journal={IEEE Transactions on Information Forensics and Security},
  volume={17},
  pages={3350--3364},
  year={2022},
  publisher={IEEE}
}

@book{hu2019eeg,
  title={EEG signal processing and feature extraction},
  author={Hu, Li and Zhang, Zhiguo},
  year={2019},
  publisher={Springer}
}

@article{yang2017usability,
  title={On the usability of electroencephalographic signals for biometric recognition: A survey},
  author={Yang, Su and Deravi, Farzin},
  journal={IEEE Transactions on Human-Machine Systems},
  volume={47},
  number={6},
  pages={958--969},
  year={2017},
  publisher={IEEE}
}

@article{marcel2007person,
  title={Person authentication using brainwaves (EEG) and maximum a posteriori model adaptation},
  author={Marcel, Sebastien and Mill{\'a}n, Jos{\'e} del R},
  journal={IEEE Trans. on pattern analysis and machine intelligence},
  volume={29},
  number={4},
  pages={743--752},
  year={2007},
  publisher={IEEE}
}

@inproceedings{kaliraman2019use,
  title={Use of EEG as a unique human biometric trait for authentication of an individual},
  author={Kaliraman, Bhawna and Singh, Priyanka and Duhan, Manoj},
  booktitle={International Conference on Advanced Communication and Computational Technology},
  pages={277--286},
  year={2019},
  organization={Springer}
}

@article{gui2019survey,
  title={A survey on brain biometrics},
  author={Gui, Qiong and Ruiz-Blondet, Maria V and Laszlo, Sarah and Jin, Zhanpeng},
  journal={ACM Computing Surveys (CSUR)},
  volume={51},
  number={6},
  pages={1--38},
  year={2019},
  publisher={ACM New York, NY, USA}
}

@article{rui2018survey,
  title={A survey on biometric authentication: Toward secure and privacy-preserving identification},
  author={Rui, Zhang and Yan, Zheng},
  journal={IEEE access},
  volume={7},
  pages={5994--6009},
  year={2018},
  publisher={IEEE}
}

@article{gupta2022review,
  title={A review of different ECG classification/detection techniques for improved medical applications},
  author={Gupta, Varun and Saxena, Nitin Kumar and Kanungo, Abhas and Gupta, Anmol and Kumar, Parvin and Salim},
  journal={Int. Journal of System Assurance Engineering and Management},
  volume={13},
  number={3},
  pages={1037--1051},
  year={2022},
  publisher={Springer}
}

@article{kawala2021summary,
  title={Summary of over fifty years with brain-computer interfaces—a review},
  author={Kawala-Sterniuk, Aleksandra and Browarska, Natalia and Al-Bakri, Amir and Pelc, Mariusz and Zygarlicki, Jaroslaw and Sidikova, Michaela and Martinek, Radek and Gorzelanczyk, Edward Jacek},
  journal={Brain Sciences},
  volume={11},
  number={1},
  pages={43},
  year={2021},
  publisher={MDPI}
}

@article{das2016eeg,
  title={EEG biometrics using visual stimuli: A longitudinal study},
  author={Das, Rig and Maiorana, Emanuele and Campisi, Patrizio},
  journal={IEEE Signal Processing Letters},
  volume={23},
  number={3},
  pages={341--345},
  year={2016},
  publisher={IEEE}
}

@article{huang2022m3cv,
  title={M3CV: A multi-subject, multi-session, and multi-task database for EEG-based biometrics challenge},
  author={Huang, Gan and Hu, Zhenxing and Chen, Weize and Zhang, Shaorong and Liang, Zhen and Li, Linling and Zhang, Li and Zhang, Zhiguo},
  journal={NeuroImage},
  volume={264},
  pages={119666},
  year={2022},
  publisher={Elsevier}
}

@inproceedings{schons2018convolutional,
  author={Schons, Thiago and Moreira, Gladston JP and Silva, Pedro HL and Coelho, Vitor N and Luz, Eduardo JS},
  booktitle={Progress in Pattern Recognition, Image Analysis, Computer Vision, and Applications: 22nd Iberoamerican Congress, CIARP 2017, Valparaiso, Chile, November 7--10, 2017, Proceedings 22},
  pages={601--608},
  year={2018},
  organization={Springer}
}

@article{bidgoly2022towards,
  title={Towards a universal and privacy preserving EEG-based authentication system},
  author={Bidgoly, Amir Jalaly and Bidgoly, Hamed Jalaly and Arezoumand, Zeynab},
  journal={Scientific Reports},
  volume={12},
  number={1},
  pages={2531},
  year={2022},
  publisher={Nature Publishing Group UK London}
}

@inproceedings{fallahi2023brainnet,
  title={BrainNet: Improving Brainwave-based Biometric Recognition with Siamese Networks},
  author={Fallahi, Matin and Strufe, Thorsten and Arias-Cabarcos, Patricia},
  booktitle={2023 IEEE International Conference on Pervasive Computing and Communications (PerCom)},
  pages={53--60},
  year={2023},
  organization={IEEE}
}

@article{maiorana2017longitudinal,
  title={Longitudinal evaluation of EEG-based biometric recognition},
  author={Maiorana, Emanuele and Campisi, Patrizio},
  journal={IEEE transactions on Information Forensics and Security},
  volume={13},
  number={5},
  pages={1123--1138},
  year={2017},
  publisher={IEEE}
}

@article{arias2023performance,
  title={Performance and usability evaluation of brainwave authentication techniques with consumer devices},
  author={Arias-Cabarcos, Patricia and Fallahi, Matin and Habrich, Thilo and Schulze, Karen and Becker, Christian and Strufe, Thorsten},
  journal={ACM Transactions on Privacy and Security},
  volume={26},
  number={3},
  pages={1--36},
  year={2023},
  publisher={ACM New York, NY}
}

@inproceedings{nakanishi2015brain,
  title={Brain waves as unconscious biometrics towards continuous authentication-the effects of introducing PCA into feature extraction},
  author={Nakanishi, Isao and Yoshikawa, Takuya},
  booktitle={2015 International Symposium on Intelligent Signal Processing and Communication Systems (ISPACS)},
  pages={422--425},
  year={2015},
  organization={IEEE}
}

@inproceedings{lin2018brain,
  title={Brain password: A secure and truly cancelable brain biometrics for smart headwear},
  author={Lin, Feng and Cho, Kun Woo and Song, Chen and Xu, Wenyao and Jin, Zhanpeng},
  booktitle={Proceedings of the 16th Annual International Conference on Mobile Systems, Applications, and Services},
  pages={296--309},
  year={2018}
}

@article{fallahi2024beyond,
  title={Beyond Gaze Points: Augmenting Eye Movement with Brainwave Data for Multimodal User Authentication in Extended Reality},
  author={Fallahi, Matin and Arias-Cabarcos, Patricia and Strufe, Thorsten},
  journal={arXiv preprint arXiv:2404.18694},
  year={2024}
}

@article{chaurasia2024neuroidbench, title={NeuroIDBench: An open-source benchmark framework for the standardization of methodology in brainwave-based authentication research}, author={Chaurasia, Avinash Kumar and Fallahi, Matin and Strufe, Thorsten and Terh{\"o}rst, Philipp and Cabarcos, Patricia Arias}, journal={Journal of Information Security and Applications}, volume={85}, pages={103832}, year={2024}, publisher={Elsevier} }

@article{wu2018eeg,
  title={An EEG-based person authentication system with open-set capability combining eye blinking signals},
  author={Wu, Qunjian and Zeng, Ying and Zhang, Chi and Tong, Li and Yan, Bin},
  journal={Sensors},
  volume={18},
  number={2},
  pages={335},
  year={2018},
  publisher={MDPI}
}

@article{maiorana2021learning,
  title={Learning deep features for task-independent EEG-based biometric verification},
  author={Maiorana, Emanuele and},
  journal={Pattern Recognition Letters},
  volume={143},
  pages={122--129},
  year={2021},
  publisher={Elsevier}
}

@article{friedman2022biometric,
  title={Biometric performance as a function of gallery size},
  author={Friedman, Lee and Stern, Hal and Prokopenko, Vladyslav and Djanian, Shagen and Griffith, Henry and Komogortsev, Oleg},
  journal={Applied Sciences},
  volume={12},
  number={21},
  pages={11144},
  year={2022},
  publisher={MDPI}
}

@misc{beyondreality2022,
  title        = {Beyond Reality: Is the long-awaited VR revolution finally on the horizon?},
  author       = {Fergus Navaratnam-Blair and Keith Wagstaff and Grady Miller and Marlon Cumberbatch and Chris Rethore},
  organization = {National Research Group},
  year         = {2022},
  month        = {April},
  note         = {Design by Grace Stees, Illustrations by Hannah Robinson},
  note         = {Accessed: 2024-12-16}
}

\appendices

\section{Session Count Distribution}

\begin{figure}[ht!]
    \centering
    \includegraphics[width=0.7\linewidth]{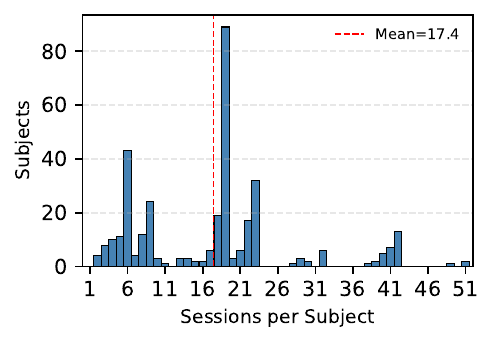}
    \caption{Histogram of session counts per subject.}
    \label{fig:session_distribution}
\end{figure}

Figure~\ref{fig:session_distribution} shows the session count distribution in our dataset.

\section{Architectural Details of Deep Learning Feature Encoders}
\label{A:B}

This appendix provides detailed specifications for the Convolutional Neural Network (CNN) architectures employed as feature encoders in our benchmarking study. In our experiments, CNN-based models trained with Supervised Contrastive Loss (SupConLoss)~\cite{khosla2020supervised} generally achieved the best performance among the evaluated feature extraction methods, with the strongest results obtained using ResNet1D.

\subsection{Best-Performing Architecture: ResNet1D}
The core concept of the Residual Network was introduced by He et al.~\cite{he2016deep} to address the degradation problem in very deep neural networks. By incorporating skip connections (residual blocks), ResNet allows information to bypass one or more layers, enabling the stable training of networks with significantly greater depth. In our work, we adopt the ResNet1D backbone proposed by Zheng et al.~\cite{zheng2022task} for EEG anomaly detection, which is based on a ResNet-34-style architecture with temporally oriented convolutional kernels for multi-channel EEG segments.

\subsubsection*{Architecture Detail}
ResNet1D is a deep CNN built from residual blocks that operate along the temporal dimension of the EEG. An initial temporal convolutional block is followed by several stages of stacked residual blocks. Each residual block contains one or more 1D convolutional layers with batch normalization (BN) and rectified linear unit (ReLU) activations, together with a skip connection from input to output. After the last residual stage, global average pooling over time is applied and the resulting feature vector is passed through a fully connected layer, which provides the final feature embedding for comparison.

\subsection{ShallowConvNet}
Introduced by Schirrmeister et al.~\cite{schirrmeister2017deep}, the ShallowConvNet was designed to closely mimic the successful Filter Bank Common Spatial Pattern (FBCSP) pipeline~\cite{ang2008filter} in an end-to-end deep learning framework. The goal was to demonstrate that a simple deep-learning structure could perform as well.

\subsubsection*{Architecture Detail}
This architecture consists of two convolutional layers, specifically engineered to extract frequency and spatial information, followed by operations that implement log-variance features:
\begin{enumerate}
    \item Temporal filtering (Block 1): A convolution with kernel size $1 \times 25$ is applied along the temporal dimension. This layer learns temporal filters independently across each EEG channel.
    \item Spatial filtering (Block 2): A convolution with kernel size $Channels \times 1$ is used to linearly combine the temporal features across electrodes, followed by batch normalization (BN).
    \item Feature extraction: The output of the spatial filter then passes through a squaring operation ($x \mapsto x^2$), followed by mean pooling (size $1 \times 75$, stride $15$), and a logarithmic activation ($\log(x)$). These operations directly implement the log-variance feature computation analogous to FBCSP.
\end{enumerate}

\subsection{DeepConvNet}
Also introduced by Schirrmeister et al.~\cite{schirrmeister2017deep}, DeepConvNet was designed as a more generic CNN architecture, closer to models used in computer vision such as VGG~\cite{simonyan2014very}. The motivation was to explore whether less specialized, deeper networks could capture a wider variety of abstract features from the raw EEG signal, rather than being biased towards the amplitude-based features commonly used in specific task decoding.

\subsubsection*{Architecture Detail}
The network is constructed with four sequential blocks, using smaller kernels and max pooling for feature compression:
\begin{enumerate}
    \item Block 1: Temporal convolution (size $1 \times 10$) followed by a spatial convolution (size $Channels \times 1$), an exponential linear unit (ELU) non-linearity, and max pooling (size $1 \times 3$, stride $3$).
    \item Blocks 2, 3, 4: These three blocks are stacked consecutively. Each consists of a convolution (size $1 \times 10$), an ELU activation, and max pooling (size $1 \times 3$, stride $3$). These blocks use small convolutional kernels to extract progressively abstract and compressed features.
    \item Output: The final output is flattened and passed through a dense (fully connected) layer, which provides the feature embedding for comparison.
\end{enumerate}

\subsection{EEGNet}
Lawhern et al.~\cite{lawhern2018eegnet} introduced EEGNet to create a compact and generalized CNN that could perform well across multiple brain-computer interface (BCI) paradigms (e.g., P300, SMR, ERN) and datasets. The primary design constraint was to achieve this with a minimal number of trainable parameters, making the model suitable for deployment in resource-constrained or real-time BCI systems.

\subsubsection*{Architecture Detail}
EEGNet achieves its efficiency through the factorization of standard convolutions into depthwise and separable convolutions:
\begin{enumerate}
    \item Block 1 (temporal–spatial filter): This block consists of a temporal convolution followed by a depthwise convolution. Temporal filters learn frequency features. The depthwise filters then apply a unique spatial filter to the output of each temporal filter, extracting frequency-specific spatial patterns.
    \item Block 2 (feature mixing): This block employs a separable convolution, which is the combination of a depthwise convolution (providing a temporal summary for each feature map) and a pointwise convolution (a $1 \times 1$ convolution). The pointwise convolution acts as an efficient final layer, learning how to optimally combine or mix the feature maps across channels.
\end{enumerate}

\begin{IEEEbiography}[{\includegraphics[width=1in,height=1.25in,clip,keepaspectratio]{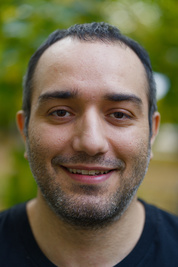}}]{Matin Fallahi}
is currently pursuing the Ph.D. degree in computer science with the KASTEL -- Institute of Information Security and Dependability, Karlsruhe Institute of Technology (KIT), Karlsruhe, Germany. The research focuses on biometric authentication, in particular brainwave-based user authentication using electroencephalography (EEG). This includes the security and robustness of EEG-based biometrics, usability and continuous authentication in extended reality (XR) scenarios, and privacy-preserving processing of biometric data. Fallahi has coauthored several publications on brainwave-based biometrics and related topics in usable security and privacy.
\end{IEEEbiography}

\begin{IEEEbiography}[{\includegraphics[width=1in,height=1.25in,clip,keepaspectratio]{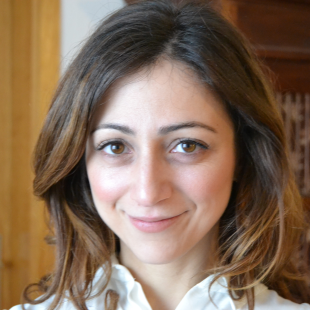}}]{Patricia Arias-Cabarcos}
is a Professor of IT Security at the University of Paderborn, where she leads the Human-centered IT Security (HITS) group. Her research focuses on usable authentication, behavioral data privacy, data transparency, and related topics in usable security and privacy. Before joining Paderborn, she was a Senior Researcher at the Karlsruhe Institute of Technology (KIT) from 2019 to 2021, a Humboldt Fellow at the University of Mannheim between 2017 and 2019, and an Assistant Professor at Universidad Carlos III de Madrid (2013–2018), where she also obtained her Ph.D. in Telematic Engineering. She has additionally held positions as a visiting postdoctoral researcher at TU Darmstadt and as a research intern at NEC Laboratories Europe in Heidelberg.
\end{IEEEbiography}

\begin{IEEEbiography}[{\includegraphics[width=1in,height=1.25in,clip,keepaspectratio]{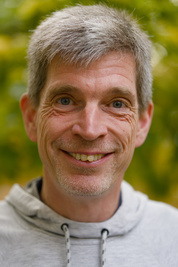}}]
{Thorsten Strufe}
is Professor of IT Security at KIT, adjunct professor for Privacy and Network Security at TU Dresden, Deputy Speaker of the Cluster of Excellence CeTI (Centre for Tactile Internet with Human-in-the-Loop), and PI in the German Competence Center for IT Security KASTEL, the Research Training Group RoSI (TU Dresden), and the 5G-Lab Germany. Prior to his position at KIT, he was full professor of privacy and security at TU Dresden from 2014 to 2019, received an offer for the Chair of Security and Privacy at University of Freiburg, was assistant professor for Peer-to-Peer Networks at TU Darmstadt from 2009 to 2014, and deputy professor for Secure Distributed Systems at the University of Mannheim throughout 2011. During this time, he was PI in several research groups and collaborative projects, such as the SFB 912 HAEC (Highly Adaptive Energy-Efficient Communication, TU Dresden), the SFB 1053 MAKI (Multi-Mechanism Adaptation for the Future Internet, TU Darmstadt), the research group 733 QuaP2P (Improving the Quality of Peer-to-Peer Systems, TU Darmstadt) and various collaborative projects within the funding programmes of EU and BMBF.
\end{IEEEbiography}

\end{document}